%% file: main.tex
\documentclass[preprint,twocolumn]{aastex701}

\newcommand{\mast}{M$_{\ast}$}
\newcommand{\msol}{M$_{\odot}$}
\newcommand{\deltasfr}{$\Delta\log(\mathrm{SFR}/\mathrm{M}_\odot \, \mathrm{yr}^{-1})$}
\newcommand{\mdeltasfr}{M$_{\ast}$-$\Delta$SFR}
\newcommand{\logmass}{$\log(\mathrm{M}_\ast/\mathrm{M}_\odot)$}
\newcommand{\logsfr}{$\log(\mathrm{SFR}/\mathrm{M}_\odot \, \mathrm{yr}^{-1})$}
\newcommand{\metallicity}{$12+\log(\mathrm{O/H})$}
\newcommand{\deltametal}{$\Delta\log(\mathrm{O/H})$}
\newcommand{\ebvgas}{E(B-V)$_{\rm gas}$}

\newcommand{\oiii}{[O\thinspace{\sc iii}]}
\newcommand{\neiii}{[Ne\thinspace{\sc iii}]}
\newcommand{\oii}{[O\thinspace{\sc ii}]}

\newcommand{\nii}{[N\thinspace{\sc ii}]}

\newcommand{\halpha}{H$\alpha$}
\newcommand{\hbeta}{H$\beta$}

\newcommand{\oiiilam}{[O\thinspace{\sc iii}]$\lambda5008$}
\newcommand{\oiilam}{[O\thinspace{\sc ii}]$\lambda\lambda3727,3730$}
\newcommand{\niilam}{[N\thinspace{\sc ii}]$\lambda6585$}

\defcitealias{Sanders2021}{S21}
\defcitealias{Sanders2025}{S25}
\defcitealias{Clarke2025}{C25}

\begin{document}

\title{The JADES Mass-Metallicity and Fundamental Metallicity Relations at $z\gtrsim2$ Using New High-Redshift Metallicity Calibrations}

\author[0000-0001-9489-3791]{Natalie Lam}
\altaffiliation{NSF Graduate Research Fellow}
\affiliation{University of California, Los Angeles, 475 Portola Plaza, Los Angeles, CA, 90095, USA}
\email[show]{natalielamwy@astro.ucla.edu}

\author[0000-0003-1249-6392]{Leonardo Clarke}
\affiliation{University of California, Los Angeles, 475 Portola Plaza, Los Angeles, CA, 90095, USA}
\email{leoclarke@astro.ucla.edu}

\author[0000-0003-3509-4855]{Alice E. Shapley}
\affiliation{University of California, Los Angeles, 475 Portola Plaza, Los Angeles, CA, 90095, USA}
\email{aes@astro.ucla.edu}

\author[0000-0003-4792-9119]{Ryan L. Sanders}
\affiliation{University of Kentucky, 506 Library Drive, Lexington, KY, 40506, USA}
\email{ryan.sanders@uky.edu}

\author[0000-0001-8426-1141]{Michael W. Topping}
\affiliation{Steward Observatory, University of Arizona, 933 N Cherry Avenue, Tucson, AZ 85721, USA}
\email{michaeltopping@arizona.edu}

\author[0000-0003-2680-005X]{Gabriel B. Brammer}
\affiliation{Cosmic Dawn Center (DAWN), Denmark}
\affiliation{Niels Bohr Institute, University of Copenhagen, Jagtvej 128, DK-2200 Copenhagen N, Denmark}
\email{gabriel.brammer@nbi.ku.dk}

\author[0000-0001-9687-4973]{Naveen A. Reddy}
\affiliation{Department of Physics \& Astronomy, University of California, Riverside, 900 University Avenue, Riverside, CA 92521, USA}
\email{naveenr@ucr.edu}

\author[0009-0002-6186-0293]{Shreya Karthikeyan}
\affiliation{University of California, Los Angeles, 475 Portola Plaza, Los Angeles, CA, 90095, USA}
\email{skarthik@g.ucla.edu}

\begin{abstract}

We present measurements of the mass–metallicity relation (MZR) and fundamental metallicity relation (FMR) at $1.4<z<7.0$ using stacked JWST/NIRSpec spectra of 601 star-forming galaxies from the JWST Advanced Deep Extragalactic Survey (JADES). Using the most up-to-date strong-line metallicity calibrations based on high-redshift galaxies, we derive gas-phase metallicities from composite spectra binned by stellar mass, redshift, and star-forming main sequence (SFMS) offset. We find that the MZR slope evolves weakly from $z\sim0$ out to $z\sim5$, with $\gamma\sim0.21\pm0.03$, while the normalization decreases smoothly with redshift at a rate of $d\log(\mathrm{O/H})/dz\sim-0.1$ out to $z\sim4$. Beyond $z\gtrsim5$, the low-mass end continues to decline in metallicity while the high-mass end remains broadly consistent with lower-redshift relations, producing a steeper overall MZR. We additionally find evidence for a shallow anti-correlation between deviations from the MZR and SFMS at fixed stellar mass at $z\sim1.4-5$. This anti-correlation, albeit with weaker SFR coupling than observed locally, suggests that an FMR is already beginning to emerge by $z\sim5$.  Comparisons with recent observations and cosmological simulations show broad agreement, though no single simulation simultaneously reproduces the observed slopes and normalizations across all redshifts. Our results support a picture in which bursty star formation and strong stellar feedback increasingly shape the regulation of galaxy growth at high redshift, while also highlighting the need for substantially larger spectroscopic samples to robustly constrain the evolution of galaxy scaling relations at high-redshift.

\end{abstract}


\section{Introduction} \label{sec:intro}

A description of the chemical evolution of galaxies across cosmic time is crucial to our understanding of galaxy formation and growth. The gas-phase oxygen abundance of a galaxy is sensitive to several key processes such as nucleosynthesis via stars, the release of metals into the interstellar medium (ISM) in supernovae and stellar winds, the outflow of metals via galactic superwinds, and accretion of both pristine and enriched material from the intergalactic and circumgalactic media into the ISM \citep[e.g.,][]{2017MNRAS.467..115D,2017ARA&A..55..389T}. 

A powerful tracer of galaxy chemical evolution is the mass-metallicity relation (MZR), relating the gas-phase oxygen abundance of a galaxy to its stellar mass. It is well-established that these two galaxy properties are tightly correlated in the local universe, highlighting the role of galaxy halo mass in regulating the retention or escape of metals from the ISM \citep[e.g.,][]{1979A&A....80..155L,2004ApJ...613..898T}. 
At high redshifts, and at stellar masses below $\sim10^{10.5}$~\msol{}, above which the $z\sim0$ MZR asymptotes, the MZR is characterized by a power law. The normalization of the MZR in this lower-mass regime has been observed to decrease significantly between $z\sim0$ and $z\sim3$ \citep[e.g.,][]{Sanders2021,Li2023,He2024}, and this decreasing evolution appears to slow down beyond $z\gtrsim3$ \citep[e.g.,][]{Nakajima2023,Curti2024}. On the other hand, the slope of the MZR stays relatively constant out to $z\sim3$ \citep[e.g.,][]{Sanders2021}. In other regimes, however, there is significant ambiguity on the behavior of the MZR. In some studies, the slope is seen to flatten out at even lower stellar masses \citep[e.g.,][]{Li2023,He2024} and at higher redshifts \citep[e.g.,][]{Curti2024,Kotiwale2026}, while others find a steepening slope \citep[e.g.,][]{Chemerynska2024}. Even among cosmological simulations, the predicted MZRs show widely different slopes, normalizations, and redshift evolutions \citep[e.g.,][]{Marszewski2024,Pallottini2025,McClymont2026,Garcia2024,Garcia2025}. The diversity of observed MZRs and lack of consensus among theoretical models point toward the need for a larger, more representative sample across a wider range of stellar masses and redshifts to accurately map out the trend for the full galaxy population and track its evolution at higher temporal resolution. 

Locally, an even tighter correlation is observed when incorporating galaxy star-formation rates (SFRs) into the MZR, yielding what is often referred to as the fundamental metallicity relation (FMR) \citep[e.g.,][]{Mannucci2010}. The FMR is thought to represent the quasi-equilibrium state that star-forming galaxies tend toward, as traced by the gas-phase metallicity, stellar mass, and SFR \citep[e.g.,][]{2012MNRAS.421...98D,2014MNRAS.443.3643P}. A major open question is whether this relationship holds at earlier cosmic times and, if so, when this regulatory framework becomes established in the timeline of galaxy evolution. Early studies found that galaxies remained broadly consistent with the local FMR out to cosmic noon \citep{Sanders2021}, supporting the idea of a redshift-invariant relation. However, more recent JWST-era surveys have begun to reveal systematic offsets toward lower metallicities at fixed stellar mass and SFR, beginning around $z\gtrsim3$ \citep[e.g.,][]{Stanton2026} or even $z\gtrsim6$ \citep[e.g.,][]{Nakajima2023,Curti2024}. The exact nature of the relationship between stellar mass, metallicity, and SFR is not yet well constrained at these early cosmic times.

A major limitation in studies of the MZR and FMR at early cosmic times is the absence of high-resolution spectroscopic samples that are as extensive and representative as those available in the local universe out to cosmic noon \citep[e.g.,][]{Mannucci2010,AM2013,Sanders2021,Korhonen2025}. Recent JWST surveys have dramatically expanded the accessible parameter space, probing galaxies across a wide range of stellar masses (down to $\sim10^6$~\msol\ in lensed fields; e.g., \citealt{Chemerynska2024,Hsiao2026}), environments \citep[e.g.,][]{Li2025}, and redshifts extending to $z\sim10$ \citep[e.g.,][]{Nishigaki2025,Pollock2026}. However, obtaining the signal-to-noise necessary to measure faint emission lines and derive quantities such as metallicity and SFR often biases high-redshift samples toward the brightest, most actively star-forming galaxies at fixed stellar mass. Although some surveys explicitly aim to characterize and mitigate these selection effects \citep[e.g.,][]{Lewis2025}, current samples remain relatively limited in size compared to lower-redshift surveys. The relatively small sample size makes it difficult to robustly map out the evolution of galaxy scaling relations across the full high-redshift galaxy population at fine enough stellar mass or temporal resolutions to adequately constrain models of galaxy formation and growth.

As mentioned above, the gas-phase metallicity of a galaxy is a crucial physical quantity used to study the MZR and FMR. A robust approach to measuring metallicity is to determine the emissivities of the emission lines arising from each ion species of a given chemical element as well as the Balmer lines, enabling the conversion from emission-line flux to abundance relative to hydrogen \citep[e.g.,][]{1967ApJ...150..825P,Izotov2006}. Determining these emissivities requires knowledge of the H\thinspace{\sc ii} region electron temperature ($T_e$) and density, the former of which depends on the ability to robustly measure faint auroral emission lines such as [O\thinspace{\sc iii}]$\lambda 4364$ or [O\thinspace{\sc ii}]$\lambda \lambda 7322,7332$ \citep[e.g.,][]{Izotov2006,Luridiana2015,PerezMontero2017}. This method, often referred to as the ``direct" or ``$T_e$" method, is considered the ``gold standard" in studies of high-redshift galaxies.

However, these auroral emission lines are typically a hundred times fainter than their brighter nebular counterparts (e.g., [O\thinspace{\sc iii}]$\lambda\lambda 5008,4960$, [O\thinspace{\sc ii}]$\lambda \lambda 3727,3730$), rendering the $T_e$ method an impractical approach for measuring metallicities in large, representative galaxy samples at high redshift due to the very long integration times that are required. As such, the ratios of strong, rest-optical emission lines are often used to determine metallicity in the absence of a $T_e$ measurement, where these ratios are calibrated to samples of galaxies with $T_e$-based metallicity measurements \citep[e.g.,][]{2004MNRAS.348L..59P,2017MNRAS.465.1384C,2018ApJ...859..175B,Curti2020,2022ApJS..262....3N} or photoionization models \citep[e.g.,][]{2002ApJS..142...35K,2016Ap&SS.361...61D,2019ARA&A..57..511K}. 

Prior to JWST, most common strong-line metallicity indicators were calibrated using samples of local galaxies, in which the ISM conditions at fixed nebular metallicity do not resemble those found in $z\gtrsim 1$ galaxies \citep[e.g.,][]{2014ApJ...795..165S,2015ApJ...801...88S,2017ApJ...836..164S,2020MNRAS.495.4430T,2020MNRAS.499.1652T,2020MNRAS.491.1427S,2021MNRAS.502.2600R,Clarke2026}. A recent work by \citet{Sanders2025} has compiled observations of galaxies at $1.3<z<10.6$ from the AURORA survey as well as several works in the literature to develop strong-line metallicity relations calibrated to high-redshift galaxies, enabling the robust determination of gas-phase oxygen abundances at $z\gtrsim 1$.

In this work, we use a sample of 601 star-forming galaxies from the JWST Advanced Deep Extragalactic Survey (JADES) \citep{2023arXiv230602465E,2025ApJS..277....4D} to investigate the MZR and the FMR at high redshift. This sample represents the largest spectroscopic study of main-sequence star-forming galaxies spanning $1.4<z<7$. Leveraging the statistical power of this dataset and employing updated metallicity calibrations designed for high-redshift ISM conditions, we construct high signal-to-noise composite spectra in finely spaced stellar mass bins and narrower redshift intervals than previous studies. This methodology enables us to trace the evolution of the MZR at higher temporal resolution across cosmic time and to test for the presence of an FMR for the first time at $z\gtrsim3$.

The structure of the paper is as follows: in Section~\ref{sec:methods}, we detail the observations, data reduction, and spectral fitting procedures to the JADES data, the derivation of physical quantities, and the stacking methodology used to generate composite spectra. In Section~\ref{sec:results}, we present our analysis of the MZR and FMR. In Section~\ref{sec:discussion}, we compare our results with previous studies and cosmological simulations, and discuss the implications of our observed trends on the timeline of galaxy assembly and evolution. Section~\ref{sec:conclusion} summarizes our conclusions. Throughout this paper, we assume $H_0 = 70\ \rm km\ s^{-1}\ Mpc^{-1}$, $\Omega_m = 0.3$, $\Omega_\Lambda = 0.7$, and a \citet{2003PASP..115..763C} initial-mass function. We adopt the solar abundances from \citet{2021A&A...653A.141A} in which $12+\rm \log(O/H)_\odot = 8.69$ and $\rm Z_\odot = 0.014$.

\section{Observations and Analysis} \label{sec:methods}

The data set and measurements in this work are based on DR3 of JADES \citep{2023arXiv230602465E,2025ApJS..277....4D} in the GOODS-N and GOODS-S fields. The reduction and analysis of the JADES data used in this work are presented in \citet{Clarke2025} (\citetalias{Clarke2025} hereafter), and we provide an overview in the following subsections.

\subsection{NIRCam and NIRSpec Observations} \label{subsec:methods-obs}

The JADES NIRCam observations used in this analysis were obtained from MAST\footnote{https://archive.stsci.edu/hlsp/jades}. Specifically, we utilized the photometric catalog, the contents of which are described in \citet{2023ApJS..269...16R}, \citet{2023arXiv230602465E,2023arXiv231012340E}, and \citet{2024ApJ...970...31R}. The JADES catalog contains observations in several filters including JWST/NIRCam F070W, F090W, F115W, F150W, F200W, F277W, F335M, F356W, F410M, and F444W. Additional observations were taken in the F182M, F210M, and F444W filters from the FRESCO program \citep{2023MNRAS.525.2864O}, and F182M, F210M, F430M, F460M, and F480M filters from the JEMS program \citep{2023ApJS..268...64W}. The photometric catalog also utilizes HST imaging mosaics from \citet{2013ApJS..209....6I,2019ApJS..244...16W}, adding photometry in the HST/ACS F435W, F606W, F775W, F814W, and F850LP filters as well as HST/WFC3 F105W, F125W, F140W, and F160W. From the photometric catalog, we used the \texttt{KRON\_CONV} catalog FITS extension, which consists of photometry measured from F444W point-spread function- (PSF) matched images using Kron apertures.

The description of the NIRSpec data and observing strategy can be found in \citet{2024A&A...690A.288B} and \citet{2025ApJS..277....4D}. Observations were taken in the $R\sim 100$ prism mode, the $R\sim 1000$ G140M/F070LP, G235M/F170LP, and G395M/F290LP gratings, and the $R\sim 2700$ G395H/F290LP grating. In this analysis, we utilized measurements only from the $R\sim 100$ and $R\sim 1000$ dispersers. For the prism and the G235M and G395M grating observations, the data were downloaded from the DAWN JWST Archive (DJA)\footnote{https://dawn-cph.github.io/dja/}. The data reduction procedure for the spectra in these dispersers is described in \citet{2025A&A...697A.189D} and \citet{2025A&A...693A..60H}. As described in \citetalias{Clarke2025}, the G140M observations were downloaded from the data release 3 sub-section on the JADES website\footnote{https://jades-survey.github.io/scientists/data.html}. To avoid self-subtraction of extended objects during the dithered-frame background subtraction step, we performed a custom reduction of 115 extended objects, described in \citetalias{Clarke2025}. The prism spectra were flux calibrated by scaling to match the NIRCam and HST photometry, and the emission lines were fit using the procedure in \citetalias{Clarke2025}.

\subsection{Measurements of Individual Galaxies} \label{subsec:methods-indiv}

Detailed descriptions of the spectral energy distribution (SED) and emission-line fitting procedures, along with the derivations of key quantities like SFR, are provided in \citetalias{Clarke2025}. Here, we summarize the key elements. 

\subsubsection{Stellar Population Parameters} \label{subsubsec:methods-indiv-sed}

Stellar population parameters were determined through SED fitting of the photometry and the flux-calibrated prism spectra with the {\sc prospector} code \citep{2021ApJS..254...22J}. The SEDs were fit with both stellar as well as nebular emission components from {\sc Cloudy} \citep{2013RMxAA..49..137F,2017ApJ...840...44B}. FSPS was used as the stellar population synthesis model \citep{2009ApJ...699..486C,2010ApJ...712..833C}, adopting MIST isochrones \citep{2016ApJ...823..102C} and the MILES stellar library \citep{2006MNRAS.371..703S}. The stellar metallicity and dust attenuation law were selected between 0.28 $\rm Z_\odot$ + SMC \citep{2003ApJ...594..279G} or 1.4 $\rm Z_\odot$ + \citet{2000ApJ...533..682C} 
depending on which model yielded the lower $\chi^2$ value in the SED fitting, as described in \citetalias{Clarke2025} and \citet{Topping2025}. The SEDs were fit using a non-parametric star-formation history (SFH) with a continuity prior \citep{2019ApJ...876....3L}. SFHs were divided into eight time bins, with the most recent two bins spanning 0-3 Myr and 3-10 Myr in lookback time, and the remaining six logarithmically spaced back to $z=20$. 

\subsubsection{Emission Line Fitting, Reddening Corrections, and SFR Measurements} \label{subsubsec:methods-indiv-grating}

Emission lines were fit simultaneously in both prism and grating observations to account for overlapping spectra and higher-order contamination in the medium-resolution data. The flux-calibrated prism spectra constrained the absolute line fluxes, while the higher-resolution grating spectra provided leverage on blended features (e.g., H$\alpha$ and \nii) by enforcing consistent line ratios across both datasets. Redshifts were initially estimated from the brightest lines and then fixed for the full fitting procedure. Emission lines were modeled as Gaussian profiles superimposed on a continuum derived from best-fit spectral energy distribution models, with the continuum smoothed to match the prism resolution. Line widths incorporated both instrumental resolution and intrinsic velocity dispersion.

The dust reddening $E(B-V)$ was derived from the decrements of at least two detected hydrogen recombination lines, namely Pa$\alpha$, Pa$\beta$, Pa$\gamma$, H$\alpha$, H$\beta$, and H$\gamma$. The intrinsic ratios were determined using \texttt{PyNeb} \citep{Luridiana2015} under the assumption of Case B recombination, electron temperature $T_e = 15000$~K, and electron density $n_e=100\,\mathrm{cm}^{-3}$ \citep{Osterbrock2006}. With respect to H$\beta$, the intrinsic ratios of Pa$\alpha$, Pa$\beta$, Pa$\gamma$, H$\alpha$, and H$\gamma$ are 0.30, 0.15, 0.09, 2.79, and 0.47, respectively. We also adopted a Milky Way dust attenuation curve \citep{Cardelli1989} to translate the observed ratios to $E(B-V)$. Unphysical or negative $E(B-V)$ values were truncated to zero.

Using the dust-corrected H$\alpha$ luminosity, the star formation rate was calculated as
\begin{equation}
    \log\left(\frac{\mathrm{SFR}}{M_\odot,\mathrm{yr}^{-1}}\right) = \log\left(\frac{\mathrm{L}_{\mathrm{H}\alpha}}{\mathrm{erg\ s}^{-1}}\right) + C,
\end{equation}
where $C=-41.37$ for galaxies best fit by the Calz+1.4Z${\odot}$ model and $C=-41.59$ for the SMC+0.28Z$_{\odot}$ model \citep{Reddy2022}, as described in Section~\ref{subsubsec:methods-indiv-sed}. In the main MZR sample of our analysis (to be discussed in Section~\ref{subsubsec:methods-stacks-sample}), $\sim80\%$ of objects had lower $\chi^2$ values in the SED fit under the SMC+0.28Z$_{\odot}$ model. 

All parameters, including emission-line fluxes, $E(B-V)$, and SFR, were inferred within a Bayesian framework using MCMC sampling, with physically motivated priors and convergence criteria applied to ensure robust parameter estimation.

\subsection{Composite Spectra} \label{subsec:methods-stacks}

\subsubsection{Samples} \label{subsubsec:methods-stacks-sample}

We draw our JADES parent sample from \citet{2025ApJS..277....4D}, consisting of 4,086 total objects and 1,245 galaxies at $1.4\leq z < 7.0$. In addition to the raw-data-level cuts listed in \citetalias{Clarke2025}, we first ensured that all systems had specific SFRs (sSFRs) greater than $10^{-11}\,\mathrm{yr}^{-1}$ to exclude quiescent galaxies. We then removed 210 galaxies without robust $(>3\sigma)$ detections of H$\alpha$. In addition, we removed 41 objects identified as AGN based on broad Balmer components or \niilam/H$\alpha>0.5$. An additional 10 objects were removed due to spurious features near \oiilam\ fitted as $>3\sigma$ detections, leaving a sample of 984 galaxies. 

We further isolated a subsample of 601 star-forming galaxies to create composite grating spectra for a stacking analysis by requiring wavelength coverage of the emission lines used for metallicity estimates, specifically \oii$\lambda\lambda 3727,3730$, \neiii$\lambda 3870$, H$\beta$, \oiii$\lambda 5008$, \nii$\lambda 6585$, and H$\alpha$. This requirement ensures that the emission line fluxes measured from composite spectra are representative of the galaxies contributing to each stack. We name this full sample the ``MZR sample''. Following the star-forming main sequence analysis of \citetalias{Clarke2025}, the galaxies were then divided into four redshift bins, namely $1.4\leq z < 2.7$, $2.7\leq z < 4.0$, $4.0\leq z < 5.0$, and $5.0\leq z < 7.0$. The two lower-redshift bins were defined to match those of analyses using ground-based surveys such as MOSDEF \citep[e.g.,][]{Sanders2021}, and the highest redshift bin $5.0\leq z < 7.0$ represents the merger of the $5<z<6$ and $6<z<7$ subsamples from \citetalias{Clarke2025}. As discussed in \citetalias{Clarke2025}, the full sample appears to be biased toward UV-bright, high-SFR galaxies below $10^{8.5}\,M_\odot$, while remaining broadly representative of the galaxy population above this mass limit. The sample in the mass-representative range comprises 418 galaxies. 

The median redshift in each redshift range of the MZR sample is: 
\begin{itemize}
    \item $z\sim2.08$ for $1.4\leq z < 2.7$, 
    \item $z\sim3.24$ for $2.7\leq z < 4.0$, 
    \item $z\sim4.40$ for $4.0\leq z < 5.0$, and 
    \item $z\sim5.66$ for $5.0\leq z < 7.0$.
\end{itemize}

As discussed in \citetalias{Clarke2025}, we note that our sample is characterized by a lack of UV-bright galaxies in the $4.0\leq z < 5.0$ bin at $\sim$10$^{10}\ M_\odot$ in comparison with the photometric sample presented in \citet{2024MNRAS.535.2998S}. We keep this caveat in mind when interpreting the SFR and metallicity of the highest-mass stack in the $4.0\leq z < 5.0$ range.

Analysis of the FMR requires dust attenuation-corrected SFR measurements of individual galaxies, as will be discussed in Section~\ref{subsubsec:methods-stacks-ebv_ratios}. Applying this additional filter resulted in a subsample of 469 galaxies, which we call the ``FMR subsample''. The sample size in the mass-representative range is 333.

The median redshift in each redshift range of the FMR subsample is: 
\begin{itemize}
    \item $z\sim2.08$ for $1.4\leq z < 2.7$, 
    \item $z\sim3.33$ for $2.7\leq z < 4.0$, 
    \item $z\sim4.40$ for $4.0\leq z < 5.0$, and 
    \item $z\sim5.62$ for $5.0\leq z < 7.0$.
\end{itemize}

\subsubsection{Stacking Procedure} \label{subsubsec:methods-stacks-stacking}

Our binning schemes and stacking procedure largely follow those of \citet{Sanders2018,Sanders2021}, and are also described in \citet{Karthikeyan2026}. At each redshift, galaxies with $M_\ast < 10^{8.5}\,M_\odot$ were grouped into a single stellar-mass bin, while galaxies above this representative mass limit were divided into equal-number bins by stellar mass. We refer to these composite spectra as the \mast{}-stacks.

We first processed the grating spectrum for each galaxy by scaling the G140M, G235M, and G395M grating spectra to the flux calibrated prism spectrum. We calculated scaling factors between the line fluxes measured in the prism and grating spectra for the strong emission lines \oiilam, H$\gamma$, \hbeta{}, \oiiilam, \halpha{}, \niilam{}, Pa$\gamma$, Pa$\beta$, and Pa$\alpha$ requiring that each is detected at S/N $\geq3$. The grating-wide scaling factor was determined by the inverse-variance-weighted average of the emission lines that fall in the wavelength range of the grating. The three grating spectra were then stitched together, shifted into the rest frame, converted from flux density to luminosity density, and normalized to the non-dust-corrected \halpha{} luminosity measured from the prism spectrum. 

With the processed grating spectra, we then generated the composite spectra by re-sampling each spectrum onto a common wavelength grid using the \texttt{SpectRes} package \citep{spectres} before calculating the median flux density at each wavelength. The above stacking process was repeated 100 times for each composite bin (in, e.g., stellar mass) via Monte Carlo simulations to determine the uncertainties of the bin edges and median \mast{}, as well as to generate the associated error spectrum. The 100 realizations of the composite spectrum for each bin were created by allowing galaxies to shift between mass bins according to the uncertainties on their \mast{}, bootstrap resampling within the bin to account for sample variance, and perturbing individual galaxy grating spectra by their error spectra prior to median-stacking. The resultant composite spectrum and error spectrum represent the median flux and 1-$\sigma$ standard deviation at each wavelength. This stacking process was also applied to the SED models associated with each galaxy to obtain a composite SED model spectrum for each bin.

For the FMR subsample, we repeated the stacking procedure above to generate \mast{}-stacks, additionally calculating the median SFR of each stack during the Monte Carlo simulation stage. Furthermore, we created bins in both stellar mass and offset from the SFR-\mast relation (i.e., Star Forming Main-Sequence, SFMS) at each redshift range relative to the trend reported in \citetalias{Clarke2025}. We determined the SFMS for the $5.0\leq z < 7.0$ range by applying the same methodology detailed in \citetalias{Clarke2025} for the combined set of galaxies in the $5.0<z<6.0$ and $6.0<z<7.0$ samples reported in that paper. The best-fit parameters for slope, normalization, and intrinsic scatter, respectively, are $\alpha=0.91^{+0.16}_{-0.16}$, $\beta_N=0.95^{+0.05}_{-0.05}$, $\sigma_{\mathrm{int}}=0.28^{+0.04}_{-0.04}$. In each redshift range, we first divided the distribution of galaxies in half, above and below the median stellar mass. We then further split these two groups into bins of positive and negative offsets from the SFMS of that redshift, resulting in a total of four bins in each redshift range. We call this binning scheme the \mdeltasfr{}-stacks.

\begin{figure}[ht!]
    \centering
    \includegraphics[width=0.45\textwidth]{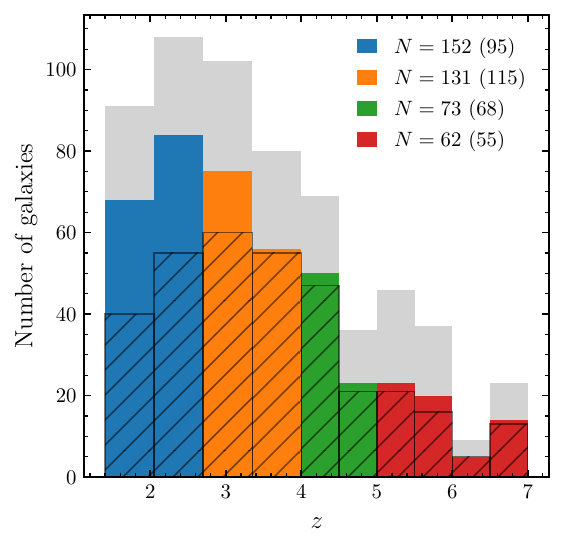}
    \caption{Redshift distribution of our sample. The full MZR sample is shown by gray bars and the colored bars highlight the sample in the mass-representative range (i.e., \logmass{}$>8.5$), with $1.4\leq z < 2.7$ in blue, $2.7\leq z < 4.0$ in orange, $4.0\leq z < 5.0$ in green, and $5.0\leq z < 7.0$ in red. The hatched bars show the distribution of galaxies in the mass-representative range of the FMR sample (i.e., galaxies with individual SFR measurements). The MZR sample numbers in the mass-representative range are displayed in the upper right, followed by those of the FMR sample in parentheses. 
    \label{fig:redshift_hist}}
\end{figure}

Figure~\ref{fig:redshift_hist} shows the redshift distributions of the MZR and FMR samples. 

\subsubsection{Emission Line Fitting} \label{subsubsec:methods-stacks-fitting}

Emission-line fluxes were measured from single-Gaussian fits to the stacked spectra. Closely spaced emission lines (within 30~\AA) and doublets with known ratios were fit simultaneously, where their centroids and width were kinematically tied. We also fit \neiii$\lambda 3870$ and the blended H I (H8) and He I feature at 3890\AA\ simultaneously.

In general, we fit linear models to the underlying continua in windows of roughly 20~\AA\ to 100~\AA\ blue- and redwards of the emission lines. For Hydrogen recombination lines and nearby emission lines, we accounted for stellar absorption using the composite SED model spectra associated with each stack. We scaled the SED model to the composite spectrum by determining a linear scaling function in continuum windows around the emission lines. The uncertainties on line fluxes were determined through Monte Carlo simulations where the composite spectra were perturbed by the error spectra 500 times.

\subsubsection{Reddening Correction, Line Ratios, and SFRs} \label{subsubsec:methods-stacks-ebv_ratios}

With the emission line fits, we calculated line ratios defined as follows:
\begin{enumerate}
    \item O3: $\log_{10}($\oiiilam/\hbeta{}$)$
    \item O32$^*$: $\rm \log_{10}\left (\frac{[O\thinspace III] \lambda5008}{[O\thinspace II]\lambda\lambda 3727, 3730} \right)$
    \item Ne3O2: $\rm \log_{10}\left (\frac{[Ne\thinspace III]\lambda 3870}{[O\thinspace II]\lambda\lambda3727,3730}\right)$  
    \item R23$^*$: $\rm \log_{10}\left (\frac{[O\thinspace III]\lambda\lambda 4960,5008 + [O\thinspace II]\lambda\lambda 3727,3730}{H\beta}\right)$
    \item N2: $\log_{10}($\niilam/\halpha{}$)$
\end{enumerate}
where line ratios with an asterisk (*) were corrected for dust attenuation. 

We first generated 500 Monte Carlo perturbations of all line fluxes based on the uncertainty on their fit. For each iteration, we calculated $E(B-V)$ exclusively from the Balmer decrement of \halpha/\hbeta of the stacked fluxes, assuming the same ISM conditions and dust law in Section~\ref{subsubsec:methods-indiv-grating} (i.e., \halpha/\hbeta~$=2.79$). The line ratios listed above were then computed, with dust-correction applied where relevant. The final value and uncertainty of each line ratio and \ebvgas{} are reported as the median and 1-$\sigma$ standard deviation across the Monte Carlo samples. 

As noted in Section~\ref{subsubsec:methods-stacks-stacking}, the SFRs of the FMR \mast{} and \mdeltasfr{} stacks were defined as the median SFRs of their constituent galaxies. This median SFR and its uncertainty was determined as part of the Monte Carlo simulations during the stacking process in Section~\ref{subsubsec:methods-stacks-stacking}, where the individual SFRs in a bin were perturbed by their uncertainties in each of the 100 realizations. 

\begin{figure*}[ht!]
    \centering
    \includegraphics[width=0.9\textwidth]{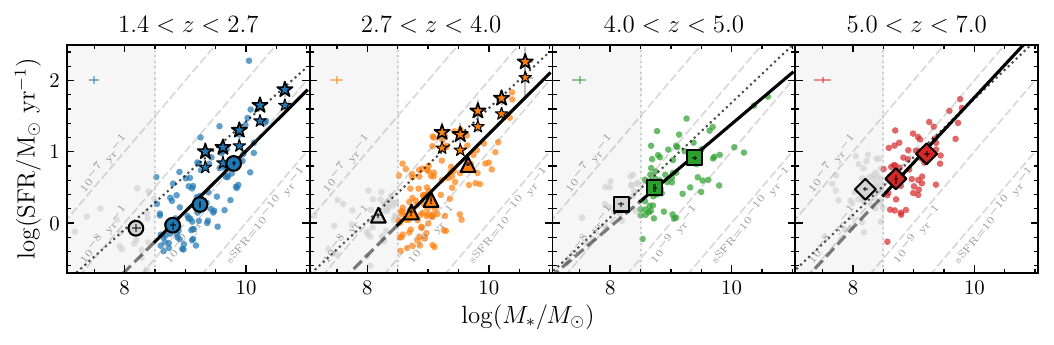}
    \caption{SFR vs. \mast{} in bins of redshift. Because not all galaxies in the full MZR sample have individual SFR measurements, the background scatters show galaxies in the FMR sample and the median of their uncertainties are denoted in the upper left corner. The larger markers represent the corresponding FMR \mast{} stacks, with the stacked SFR values determined by the median SFR of the galaxies that contribute to each stack. Points with \logmass$<8.5$ are outside of the mass-representative range and are colored gray. The $z\sim2.3$ and $z\sim3.3$ stacked points from \citetalias{Sanders2021} are displayed as stars in the $1.4\leq z < 2.7$ and $2.7\leq z < 4.0$ panels. The smaller, fainter star markers show these same stacks, with the SFR adjusted by $-0.32$ dex to match the low-metallicity H$\alpha$ to SFR conversion that we adopt for the majority of the objects in this work. The uncertainties on the star markers are typically smaller than the marker size. The solid black lines in the left three panels denote the SFMS fit from \citetalias{Clarke2025} and the fit in the rightmost panel is performed using the same methodology and based on galaxies in the $5.0\leq z < 7.0$ range reported in that paper. The black dotted lines indicate the SFMS fit from \citet{Speagle2014} for $z=$ 2.08, 3.33, 4.40, and 5.62 for each panel, respectively.
    \label{fig:SFMS}}
\end{figure*}

Figure~\ref{fig:SFMS} presents the SFRs of individual galaxies and the stacks vs. \mast{}. Our FMR sample, which is a subsample of that of \citetalias{Clarke2025}, follows the best-fit SFMS well and indicates that there are no systematic biases in our subsample. For comparison, we also show the stacked MOSDEF points from \citet{Sanders2021} (\citetalias{Sanders2021} hereafter) on the SFMS (large, bold outlined stars) as well as these same stacks offset by $-0.32$ dex to match the low-metallicity H$\alpha$ to SFR conversion that characterizes the majority of the objects presented in this work (smaller, fainter stars).

\subsection{Metallicities} \label{subsec:methods-metal}

To robustly infer gas-phase metallicities for our sample of galaxies, we adopt the strong-line metallicity calibrations presented in \citet{Sanders2025} (\citetalias{Sanders2025} hereafter), which are based star-forming galaxies at $z\sim2-10$ from the AURORA Survey. These updated relations derived from high-redshift galaxies better reflect the evolving ISM conditions at $z\gtrsim2$, mitigating systematic biases that arise from applying the more commonly used local Universe calibrations on these early galaxies. 

Following \citetalias{Sanders2021}, we determined gas-phase metallicities by considering a combination of the strong emission line ratios O3, O32, and Ne3O2 that are detected in the spectra of individual galaxies and in the stacked spectra. These three line ratios contain emission lines of only hydrogen and $\alpha$-elements, the latter of which are produced by core-collapse supernovae and thus trace the gas-phase oxygen abundance more directly. These line ratios are therefore not susceptible to evolving relations such as the nitrogen-to-oxygen ratio (N/O) if N2 were used. Due to the degenerate nature of O3 in the typical ranges of metallicities found in our sample, we required the detection of at least one additional line ratio (O32 or Ne3O2) in order to break the degeneracy in the spectra of individual galaxies. In such spectra, a sole detection of O32 or Ne3O2 was also allowed for metallicity determination. In our stacked spectra, all three line ratios O3, O32, and Ne3O2 are available in the mass-representative range, enabling robust measurements using this full set of strong-line calibrations.

For the ratios O32 and Ne3O2, we applied the recommended quadratic parameterizations provided in \citetalias{Sanders2025}. The best-fit metallicities were calculated via $\chi^2$ minimization, and the uncertainties were determined through a Monte Carlo method by perturbing the ratios by their uncertainties 1000 times. The $\chi^2$ is defined as
\begin{equation}
    \chi^2(x) = \sum_i \frac{\left[\log R_{i,\mathrm{obs}}-\log R_{i,\mathrm{cal}}(x)\right]^2}{\sigma_{i,\mathrm{obs}}^2 + \sigma_{i,\mathrm{cal}}^2},
\end{equation}
where $x = 12+\log(\mathrm{O/H}) - 8.0$, $R_{i,\mathrm{obs}}$ is the observed value of ratio $i$, $R_{i,\mathrm{cal}}(x)$ is the modeled ratio value from the calibrations of \citetalias{Sanders2025}, $\sigma_{i,\mathrm{obs}}$ is the uncertainty of the observed ratio, and $\sigma_{i,\mathrm{cal}}$ is the intrinsic scatter in ratio values at fixed metallicity as reported in \citetalias{Sanders2025}.

The measured line ratios and best-fit metallicities for the stacked spectra are listed in Tables~\ref{tab:MZR_stacks}–\ref{tab:anticorr_stacks}. We also provide the associated median galaxy properties for each stack, including \mast{} and SFR. The tables include the \mast{}-stacks used for the MZR and FMR analyses, as well as the \mdeltasfr{}-stacks constructed from the FMR subsample.

\input{table_mzr_stacks_nosfr.tex}

\input{table_fmr_stacks.tex}

\input{table_anticorr_stacks.tex}

\section{Results} \label{sec:results}

\subsection{Trends between Line Ratios and Stellar Mass} \label{subsec:results-line_ratios}

\begin{figure*}[ht!]
    \centering
    \includegraphics[width=\textwidth]{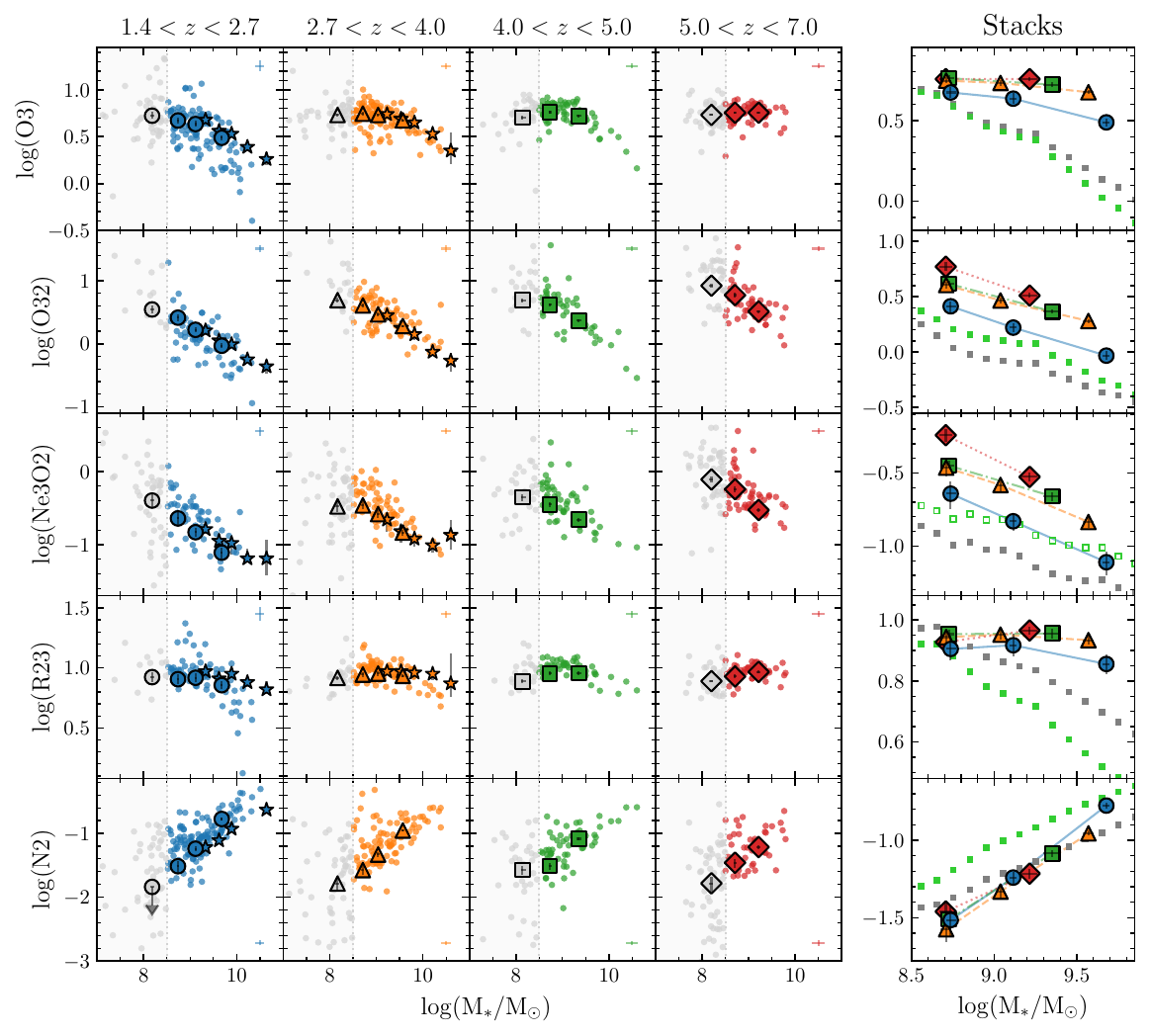}
    \caption{Emission-line ratios vs. \mast{} in bins of redshift for the MZR sample. \textit{Left:} Small circles show individual galaxy measurements with $\mathrm{S/N}>3$ detections in component lines, whereas the larger markers with black outlines represent measurements from the stacks as listed in Table~\ref{tab:MZR_stacks}. The star markers in the two lowest redshift ranges correspond to the values from $z\sim2.3$ and $z\sim3.3$ stacks to the MOSDEF survey data reported in \citetalias{Sanders2021}. Points with \logmass{}$<8.5$ are grayed out. Error bars on the stacked points are shown and the median error bars of the individual galaxies in each panel are displayed in the upper right corner (lower right corner for the N2 ratio). Arrows denote the 3-$\sigma$ upper limits. \textit{Right:} This column shows only the stacked points in the mass-representative range. The marker shapes and color schemes follow those of the left panels, and faint lines connecting points are included to guide the eye. The emission line ratios at $z\sim0$ from \citetalias{Sanders2021} and the DIG-corrected counterparts are shown as gray and lime green squares, respectively, in the background of each panel. The hollow lime green squares in the Ne3O2 panel indicate upper-limits as DIG correction is not available for [Ne\thinspace{\sc iii}].
    \label{fig:line_ratio_mosaic}}
\end{figure*}

We first investigate the empirical trends between strong emission-line ratios and \mast{} in different redshift ranges. Figure~\ref{fig:line_ratio_mosaic} shows the ratios O3, O32, Ne3O2, R23, and N2 against \mast{} for the objects in our MZR sample. The rightmost column isolates the line ratios obtained from the composite spectra, along with the $z\sim0$ trends of SDSS data \citepalias{Sanders2021}. In the lowest two redshift ranges, our stacked points agree very well with the overlaid $z\sim2.3$ and $z\sim3.3$ MOSDEF stacked line ratios from \citetalias{Sanders2021} indicated as star-shaped points. 

The scatter of O3 ratios exhibits a decreasing trend with increasing mass in the lowest redshift bin, with the trend becoming increasingly flat with \mast{} as redshift increases. We find a decreasing trend in O32 and Ne3O2 with increasing \mast{} across all redshift ranges, whereas N2 increases with increasing \mast{}. R23, on the other hand, does not show strong variations with respect to \mast{} across our redshift ranges. In general, we recover the commonly observed trends of metallicity-sensitive strong-line ratios with mass, finding that most ratios, with the exception of O3 and R23, show strong correlations with mass.

The mild trends in the O3 and R23 ratios with stellar mass can be explained by the double-valued nature of these calibrations, in which the line ratios are anti-correlated with metallicity above a characteristic ``turnover metallicity," below which the ratios are positively correlated. For the O3 and R23 calibrations, the majority of our sample lies in this range of ``turnover" metallicities, resulting in the observed flat trend of these ratios with stellar mass \citep[e.g.,][]{Steidel2014}. The turnovers in the O32 and Ne3O2 calibrations, however, occur at very low metallicities \citep[\metallicity$\lesssim7.4$; e.g., ][]{Nakajima2023,Sanders2025}, essentially behaving as monotonic relations in the metallicity range spanned by our sample. For example, some of the highest O32 and Ne3O2 ratios in the MZR sample of $\log(\mathrm{O32})=0.7$ and $\log(\mathrm{Ne3O2})=-0.3$ correspond to a metallicity value of $12+\log(\mathrm{O/H})\sim8.0$, representing some of the lowest metallicities in our sample. Coincidentally, the O3 and R23 ratios peak at \metallicity$\sim8.0$. Thus, most of our sample lies in the flat part of the O3 and R23 calibrations, with a small fraction of objects populating the high-metallicity branch where smaller ratios correspond to higher metallicities. This effect is especially evident in the R23 calibration from \citetalias{Sanders2025}, which has a wider and flatter peak. For a typical uncertainty of $\delta \log(\mathrm{R23})\sim0.05$, the metallicity inferred from the R23 calibration spans $7.5\lesssim12+\log(\mathrm{O/H})\lesssim8.4$, leading to great degeneracy.

In the rightmost column of Figure~\ref{fig:line_ratio_mosaic}, there is evident evolution of the line ratio vs. \mast{} relations between the $z\sim0$ stacks and our high-redshift stacks, as well as evolution between the different high-redshift ranges. The observed trend of increasing line ratio values (i.e., decreasing metallicities) at fixed \mast{} toward higher redshift explains the flattening of the O3- and R23-\mast{} relationships. At higher redshifts, galaxies typically exhibit lower metallicities at fixed \mast{}, resulting in their population of the turnover region of the O3 and R23 calibrations around $7.7\lesssim12+\log(\mathrm{O/H})\lesssim8.2$. The lack of constraining power from the R23 ratio, along with systematic uncertainties in the evolution of galaxy N/O abundances and ionization conditions affecting the \nii/H$\alpha$ ratio \citep[e.g., ][]{Masters2014,Runco2021}, suggest that our choice to omit these two calibrations when calculating metallicities is justified (as discussed in Section~\ref{subsec:methods-metal}).

\subsection{Mass-Metallicity Relation} \label{subsec:results-MZR}

\begin{figure*}[ht]
    \centering
    \includegraphics[width=\textwidth]{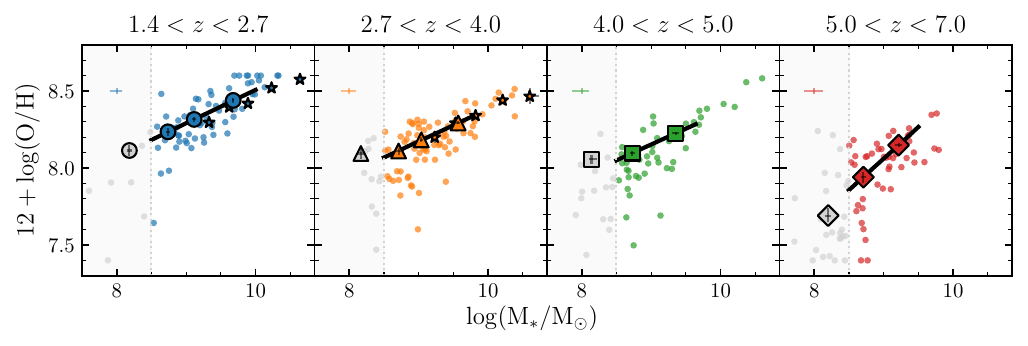}
    \caption{Best-fit metallicity vs. \mast{} in bins of redshift for the MZR sample. The distribution of individual galaxies and points follow the marker shapes and color scheme of Fig.~\ref{fig:line_ratio_mosaic}. The best-fit linear MZR to the stacked points in the mass-representative range in each redshift range is shown by a solid black line. The star markers in the left two panels indicate the $z\sim2.3$ and $z\sim3.3$ MOSDEF stacked points from \citetalias{Sanders2021} with application of the updated metallicity calibrations of \citetalias{Sanders2025}.
    \label{fig:MZR_zranges}}
\end{figure*}

In Figure~\ref{fig:MZR_zranges}, we present the best-fit metallicities vs. \mast{} at each redshift range for the MZR sample, where we recover a positive trend in all panels and a decreasing normalization in metallicity towards higher redshift. We applied the same procedure described in Section~\ref{subsec:methods-metal} to the \citetalias{Sanders2021} MOSDEF points shown by star markers. As the median redshift of the \citetalias{Sanders2021} MOSDEF $z\sim2.3$ sample is roughly 0.2 higher than our $1.4\leq z < 2.7$ sample, applying $d\log(\mathrm{O/H})/dz = -0.11$ from \citetalias{Sanders2021} shifts the metallicity of the \citetalias{Sanders2021} MOSDEF stacked points in this redshift range up by $\sim0.02$ dex. After this correction, the metallicity measurements of our $1.4\leq z<2.7$ and $2.7\leq z<4.0$ stacks are in good agreement with those of the $z\sim2.3$ and $z\sim3.3$ MOSDEF stacks from \citetalias{Sanders2021}.

We note that the distribution of galaxies in the high-mass end of the $1.4\leq z < 2.7$ sample clips at \metallicity$=8.6$, the upper bound of the range of metallicities over which the calibrations of \citetalias{Sanders2025} are valid. This effect is driven by the low O3, O32, and Ne3O2 ratios that extend below the calibration limits. The opposite is true for the low-mass end of the $5.0\leq z < 7.0$ sample where data points populate very low metallicities. These galaxies exhibit high O32 and Ne3O2 ratios that lie beyond the region covered by the quadratic parametrization of the calibrations. As a result, the minimum metallicity of the calibration range is assigned to these galaxies for O32 and Ne3O2, thus bringing down the overall best-fit metallicity values during the $\chi^2$  minimization procedure. Although the linear parameterizations of O32 and Ne3O2 extend into the regime of these high line ratios and would therefore alleviate this issue, we still adopted the formulations recommended by \citetalias{Sanders2025}, since our primary analysis focuses on stacked spectra whose line ratios do not lie within these extreme regimes.

\begin{figure}[ht]
    \centering
    \includegraphics[width=0.49\textwidth]{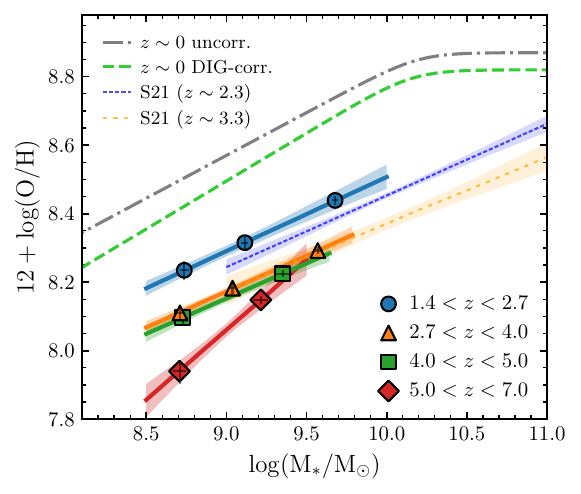}
    \caption{The MZR for stacked spectra in different redshift ranges. The points follow the marker shapes and color scheme of Fig.~\ref{fig:line_ratio_mosaic}. The solid lines represent the best-fit linear MZR in each redshift range. The blue and orange dashed lines trace the MZR fits to the $z\sim2.3$ and $z\sim3.3$ \citetalias{Sanders2021} MOSDEF stacks shown in Fig.~\ref{fig:MZR_zranges}. The shaded regions around these fitted lines denote the $1\sigma$ uncertainties of the fits. The $z\sim0$ fits to the DIG-corrected (green) and uncorrected (gray) SDSS data from \citetalias{Sanders2021} are also overlaid. The slope of the MZR holds out to $z\sim5$ and steepens beyond, while the normalization of the MZR decreases in metallicity at fixed mass towards earlier cosmic times. 
    \label{fig:MZR_summary}}
\end{figure}

Considering that our stacked points all correspond to median masses lower than the typical MZR turnover mass of \mast$\sim10^{10}\,M_\odot$, we model the MZR as a linear function
\begin{equation}
    12+\log(\mathrm{O/H})=\gamma\times \log\left(\frac{M_\ast}{10^9\,M_\odot}\right) + Z_9
\end{equation}
where $\gamma$ is the slope and $Z_9$ is the metallicity at $M_\ast=10^9$~\msol{}. We performed a linear fit to the stacked points in each redshift range via a Monte Carlo simulation approach, perturbing the stellar masses and the metallicities by their uncertainties 1000 times. Following the simulations, we report the median and the standard deviation of the MZR fit parameters in Table~\ref{tab:MZR_fits}. The valid mass range for each MZR was determined as follows. To determine the upper bound of the stellar mass range we consider, we extend past the highest-mass stack by an increment that is half the difference between it and the second highest-mass stack. The bounds of the stellar mass ranges extend from $10^{8.5}$~\msol{} to $10^{10}$~\msol{}, $10^{9.8}$~\msol{}, $10^{9.7}$~\msol{}, and $10^{9.5}$~\msol{} for the $1.4\leq z < 2.7$, $2.7\leq z < 4.0$, $4.0\leq z < 5.0$, and $5.0\leq z < 7.0$ bins, respectively. The stacked points and fits, along with the updated fits to the MOSDEF points from \citetalias{Sanders2021}, are isolated in Figure~\ref{fig:MZR_summary}. Note that the MOSDEF fits are parameterized by $Z_{10}$ instead, and $Z_9$ can be obtained from $Z_9 = Z_{10}-\gamma$. 

All high-redshift trends show a lower normalization of metallicity at fixed \mast{} as compared to the $z\sim0$ relation. There is a $\sim0.2$~dex decrease in normalization between $z\sim0$ and $z\sim2.1$ at $10^9\,M_\odot$, and a $0.12$~dex decrease between $z\sim2.1$ and $z\sim3.3$. Both of these trends correspond to an evolution rate of $d\log(\mathrm{O/H})/dz = -0.10\pm0.02$, highly consistent with that reported in \citetalias{Sanders2021}. The evolution between $z\sim3.3$ and $z\sim4.4$ is not as evident, with a decrease of only $0.02$~dex, or $d\log(\mathrm{O/H})/dz = -0.02\pm0.01$. 

On the other hand, the slope of the MZR is highly consistent out to $z\sim5$ at $\gamma\sim0.21$, flatter than $\gamma=0.28$ of the DIG-corrected $z\sim0$ fit below the turnover mass. Notably, we measure a significant steepening of the MZR beyond $z\sim5$. This slope change results from the low-mass bin being significantly lower than the corresponding masses at lower redshifts, whereas the high-mass bin matches up well with the $z\sim3.24$ and $z\sim4.40$ MZR. However, more data are needed to confirm this trend as it is currently derived from only two stacked points.

\input{table_mzr_fits_wsanders.tex}

\subsection{Fundamental Metallicity Relation} \label{subsec:results-FMR}

\subsubsection{Comparision with $z\sim0$ FMR} \label{subsubsec:results-FMR-z0}

\begin{figure}[hb]
    \centering
    \includegraphics[width=0.49\textwidth]{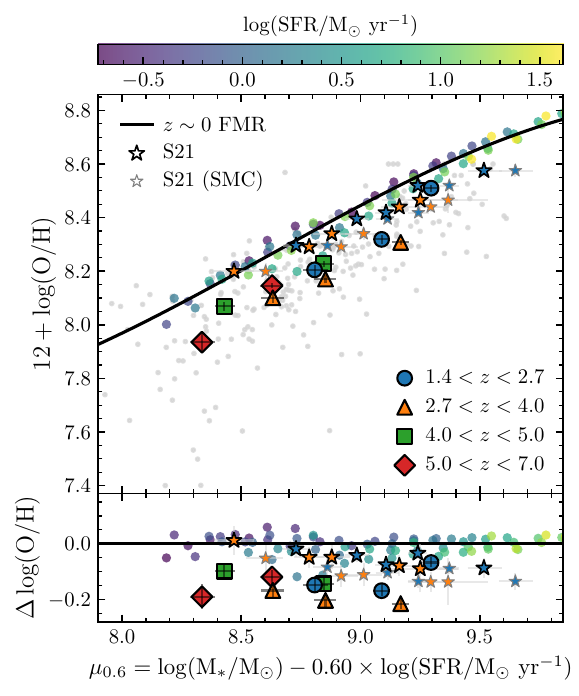}
    \caption{The FMR as shown by metallicity vs. $\mu_{0.6}=\log(\mathrm{M}_\ast/\mathrm{M}_\odot)-0.60\times\log(\mathrm{SFR}/\mathrm{M}_\odot \, \mathrm{yr}^{-1})$ (\textit{top}) and residuals in metallicity around the $z\sim0$ FMR (\textit{bottom}). The scatter of small, colored circles shows the $z\sim0$ SDSS stacks and is color-coded by SFR, while the solid black line represents the best-fit $z\sim0$ relation reported in \citetalias{Sanders2021}. The large points represent the stacks listed in Table~\ref{tab:FMR_stacks} and follow the marker shapes and color scheme of Fig.~\ref{fig:line_ratio_mosaic}. The individual galaxies in the mass-representative range of the FMR sample are denoted by the light-gray background scatter. The star markers denote the $z\sim2.3$ (blue) and $z\sim3.3$ (orange) MOSDEF stacks from \citetalias{Sanders2021} using the updated metallicity calibrations of \citetalias{Sanders2025}. The fainter, smaller star markers correspond to the MOSDEF stacks with SFR measurements based on an SMC dust attenuation law instead. 
    \label{fig:FMR_summary}}
\end{figure}

We investigate the relationship between \mast{}, metallicity, and SFR by defining the parameter $\mu_\alpha=\log(M_\ast/M_\odot)-\alpha\times\log(\mathrm{SFR}/M_\odot\,\mathrm{yr}^{-1})$ as proposed by \citet{Mannucci2010}. Figure~\ref{fig:FMR_summary} shows metallicity vs. $\mu_\alpha$ where we have set $\alpha=0.60$, which was shown to minimize the scatter of $z\sim0$ DIG-corrected SDSS stacks displayed as circles color-coded by SFR in \citetalias{Sanders2021}. The $z\sim0$ FMR is modeled with the cubic function
\begin{equation}
    12+\log(\mathrm{O/H}) = 8.80 + 0.188y - 0.220y^2 - 0.0531y^3
\end{equation}
where $y=\mu_{0.6}-10$ as defined in \citetalias{Sanders2021}. 

We display the \mast{}-stacks of the FMR subsample, as well as the \citetalias{Sanders2021} MOSDEF stacks with rederived metallicities in Figure~\ref{fig:FMR_summary}. We also indicate where these MOSDEF stacks would lie if a low-metallicity SFR conversion were applied instead of one based on \citet{Hao2011}, as we adopt for the majority of galaxies in our sample. This translation moves the MOSDEF points toward higher $\mu_{0.6}$, resulting in a more pronounced offset from the $z\sim0$ relation as compared to what was found in \citetalias{Sanders2021}. From the bottom panel of Figure~\ref{fig:FMR_summary}, which displays the residuals from the $z\sim0$ FMR, we see that all high-redshift stacks from this analysis lie below the local FMR with a typical offset of $\sim-0.15$~dex. The presence of a defined and offset locus traced by the high-redshift stacks suggests the existence of a high-redshift FMR that is offset from the $z\sim0$ relation.

\subsubsection{Anti-correlation between SFR and metallicity offsets} \label{subsubsec:results-FMR-anticorr}

It is clear that the MZR and SFMS evolve with redshift (see Section~\ref{subsec:results-MZR} and e.g., \citealt{Speagle2014}; \citetalias{Clarke2025}). An important question, therefore, is whether the underlying physical connection between star formation and chemical enrichment is already in place at early cosmic times. Rather than focusing solely on the global evolution of these scaling relations, investigating second-order trends provides insights into the physical processes that regulate fluctuations about them \citep[e.g.,][]{Dave2017,Torrey2018}. In particular, we test whether galaxies offset above the SFMS tend to exhibit systematically lower metallicities at fixed stellar mass, as expected from the local FMR.

\begin{figure*}[ht!]
    \centering
    \includegraphics[width=\textwidth]{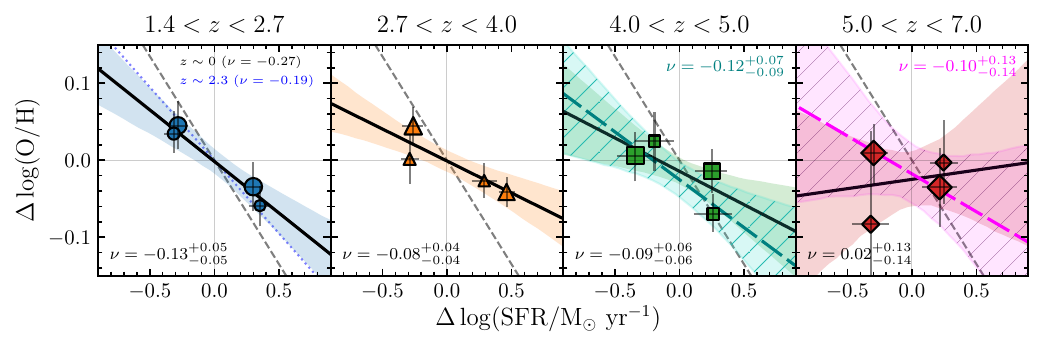}
    \caption{Metallicity residuals around the best-fit MZR ($\Delta\log(\mathrm{O/H})$) vs. SFR residuals around the SFMS ($\Delta$\logsfr) for the FMR sample in each redshift range, with bins of \mast{} and \deltasfr{} plotted. Marker sizes are varied to differentiate the lower median \mast{} stacks (smaller markers) from the higher mass stacks (larger markers). In each redshift range, the best-fit linear relation is shown by the black solid line and the shaded region denotes the $1\sigma$ uncertainty. The fitted slope $\nu$ is reported in the lower left corner of each panel. 
    In the $4.0\leq z < 5.0$ panel, the dashed green line and hatched region correspond to the best-fit relation without considering the high-mass, high-SFR point that is known to underrepresent the true high-SFR population as determined by the JADES photometric sample \citepalias{Clarke2025}. In the $5.0\leq z < 7.0$ panel, the dashed magenta line and hatched region correspond to the best-fit relation only considering the high-mass points. The fitted $\nu$ for these two additional relations are reported in the upper right corners of the respective panels.
    The black dashed line marks the $z\sim0$ relation and the blue dotted line marks the $z\sim2.3$ relation reported in \citetalias{Sanders2021}. An anti-correlation between residuals around the MZR and SFMS is seen up to $z\sim5$. 
    \label{fig:anticorr}}
\end{figure*}

With the suggestion of a potentially different FMR at high redshift based on the results in Figure~\ref{fig:FMR_summary} and Section~\ref{subsubsec:results-FMR-z0}, we investigate this relationship using the  offsets in the SFMS and MZR of the FMR \mdeltasfr{} stacks. Figure~\ref{fig:anticorr} presents the SFR deviations from the SFMS from \citetalias{Clarke2025} and Section~\ref{subsubsec:methods-stacks-ebv_ratios}, \deltasfr{}, and metallicity deviations from the fitted MZR from Section~\ref{subsec:results-MZR}, \deltametal{}, in the corresponding redshift ranges. Uncertainties on these offsets were estimated via Monte Carlo resampling, in which SFRs, metallicities, and the parameters of the SFMS and MZR were perturbed within their uncertainties over 1000 realizations. 

We fit a linear relation to these offsets within each redshift bin, with the form
\begin{equation}
    \Delta\log(\mathrm{O/H}) = \nu\times\Delta\log(\mathrm{SFR}/\mathrm{M}_\odot \, \mathrm{yr}^{-1}) + \delta_\mathrm{O/H,0}
\end{equation}
where $\nu$ is the slope and $\delta_\mathrm{O/H,0}$ is the MZR offset at $\Delta\log(\mathrm{SFR}/\mathrm{M}_\odot \, \mathrm{yr}^{-1})=0$. We allowed for a non-zero intercept because the galaxies in the FMR subsample, which have individually measured dust-corrected SFRs, are not guaranteed to span the same parameter space as the galaxies used to define our fiducial MZR, and therefore may not symmetrically scatter around it. 
Barring slight offsets in $\delta_\mathrm{O/H,0}$, this parameterization can be rewritten as $\Delta\log({\rm O/H})\propto \Delta\log({\rm SFR})^{\nu}$.

At $z\lesssim5$, we find a shallow but significant anti-correlation between metallicity and SFR, with slopes of $\nu=-0.13\pm0.05$, $\nu=-0.08\pm0.04$, and $\nu=-0.09\pm0.06$ in the $1.4\leq z < 2.7$, $2.7\leq z < 4.0$, and $4.0\leq z < 5.0$ bins, respectively. This anti-correlation is seen in both the low and high mass stacks, denoted by the smaller and larger marker sizes in Figure~\ref{fig:anticorr}. We additionally report a fit excluding the high-mass, high-SFR stack in the $4.0\leq z < 5.0$ bin, motivated by the lack of UV-bright objects in the high-mass end of this sample within this redshift range. Indeed, this high-mass, high-SFR point appears to be biased high in metallicity at fixed \deltasfr{}, which is consistent with a picture where galaxies with lower SFRs display higher metallicities at fixed mass. Removal of this point steepens the slope to $\nu=-0.12\pm0.08$, but the trend still remains shallower than that observed at $z\sim0$ of $\nu=-0.27$ \citepalias{Sanders2021}, as is the case with the slopes of the other two redshift bins. We also indicate the best-fit anti-correlation of the MOSDEF $z\sim2.3$ stacks from \citetalias{Sanders2021} for reference, where $\nu=-0.19$. We note, however, that this relation was adopted directly from that work, which employed different methodologies, metallicity calibrations, and SFR conversions than those used in our analysis; therefore, direct comparisons should be made with caution. Using the updated metallicity measurements and MZR fits to the \citetalias{Sanders2021} MOSDEF data, and under the assumption that the SFMS offsets are not affected by the constant scalar adjustment of $-0.32$~dex (see Section~\ref{subsubsec:methods-stacks-ebv_ratios}), we obtain $\nu\sim0.15$ for the $z\sim2.3$ stack, which is in closer alignment to that of our $1.4\leq z < 2.7$ bin.

In the highest redshift bin $5.0\leq z < 7.0$, we do not find clear evidence for an anti-correlation when considering all stacked points since all four points are consistent with one another in $\Delta\log({\rm O/H})$ within their errors. Isolating the high-mass points recovers a tentative anti-correlation with $\nu=-0.10\pm0.14$, though not statistically significant given the large uncertainties and limited data points. Notably, the low-mass, low-SFR stack exhibits significantly lower metallicity than all other stacks in this redshift bin, and thus drives the flattening of the best-fit line. Qualitatively, this behavior is consistent with the trend seen in the MZR in Section~\ref{subsec:results-MZR}, where the steepening of the MZR slope in the $5.0\leq z < 7.0$ bin is primarily driven by the low-mass end, while the high-mass end is seen to broadly fall in the regime spanned by the $z\sim3.3$ and $z\sim4.5$ MZR fits. However, given that each stack is composed of only $\sim13$ galaxies on average, we do not possess the numbers to robustly test this hypothesis.

Overall, we find anti-correlations between deviations from the SFMS and MZR in our three lowest-redshift bins, though the trends are weaker than those observed in the local Universe. The presence of this relation suggests that the physical processes regulating star formation, gas flows, and chemical enrichment were already becoming established by $z\sim5$.

\subsubsection{Constructing an FMR based on high-$z$ galaxies} \label{subsubsec:results-FMR-alpha}

Motivated by the presence of an anti-correlation between SFR and metallicity residuals at $z\lesssim5$, and by the systematic offset of our high-redshift sample relative to the local FMR, we applied the standard approach of optimizing the projection parameter $\alpha$ to minimize the scatter in the \mast\-Z–SFR relation \citep[e.g.,][]{Mannucci2010,AM2013,Stanton2026}. Because this analysis required both individual metallicity and individual SFR measurements, the resulting subsample of the FMR galaxies consisted of 56, 76, and 50 galaxies in the $1.4\leq z < 2.7$, $2.7\leq z < 4.0$, and $4.0\leq z < 5.0$ bins, respectively, for a total sample of 182 galaxies. To assess the representativeness of this subsample to the full FMR sample, we compared their SFR–\mast{} distributions within each redshift bin. We find that the rolling median SFRs of the subsamples are biased high by $\lesssim0.2$~dex below $\sim10^{9.5}$~\msol{} relative to the full FMR sample, while the rolling median trends agree well above $10^{9.5}$~\msol{}. 

\begin{figure*}[h!t]
    \centering
    \includegraphics[width=\textwidth]{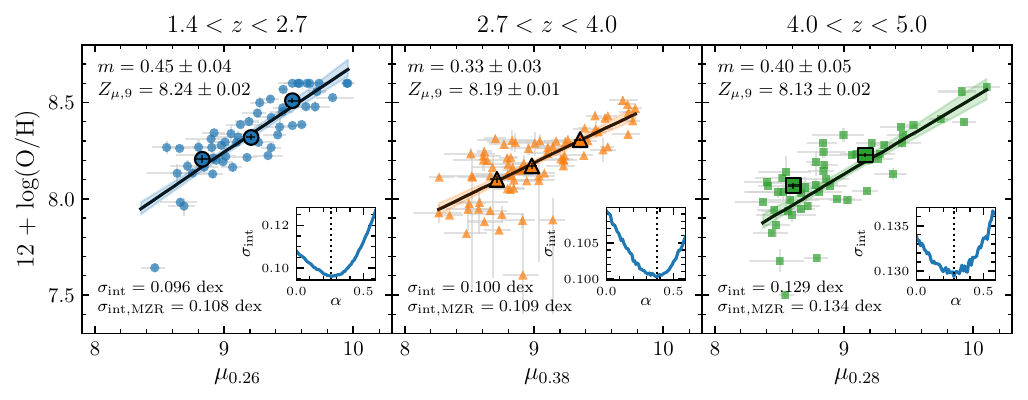}
    \caption{Tests of the FMR with projections $\mu_\alpha$ that minimize the scatter of galaxies in each redshift range. In each panel, the distribution of individual points shows the subset of component galaxies in the FMR subsample that have individual metallicity measurements, which are used to determine an $\alpha$ that minimized the intrinsic scatter $\sigma_\mathrm{int}$. The large points indicate the FMR \mast{}-stacks in the mass-representative range (i.e., with $M_\ast\geq 10^{8.5}\,M_\odot$) and are overlaid only for reference. The solid line denotes the best linear fit under the parametrization $\mu_\alpha$, the shaded region encompasses the $1\sigma$ uncertainty of the fit, and the best-fit parameters are reported in the upper left corner. The inset plot shows $\sigma_\mathrm{int}$ vs. the range of $\alpha$ sampled. We marked the best-fit $\alpha$ with the dotted line. The $\sigma_\mathrm{int}$ under the best  $\mu_\alpha$ projection and the intrinsic scatter of the MZR (i.e., when $\alpha=0$, or $\sigma_\mathrm{int,MZR}$) are reported in the lower left corner. All three redshift ranges prefer a lower value of $\alpha$ than $\alpha=0.60$ that minimizes the scatter of the SDSS stacks from \citetalias{Sanders2021}, and display mild reduction in scatter compared to the MZR. 
    \label{fig:FMR_alpha_fit}}
\end{figure*}

We constructed a grid of $\alpha$ values from 0 to 0.6 in steps of 0.01 and fit a linear relation using \texttt{linmix} \citep{linmix} to determine the best-fit parameters and intrinsic scatter ($\sigma_\mathrm{int}$) for each $\alpha$. The best-fit $\alpha$ was defined as the value that minimized this scatter. Figure~\ref{fig:FMR_alpha_fit} shows the resulting projections of the FMR for each redshift bin, with insets displaying the intrinsic scatter as a function of $\alpha$. The best-fit values of $\alpha$ are roughly $\alpha\sim0.26$ for $1.4\leq z < 2.7$, $\alpha\sim0.38$ for $2.7\leq z < 4.0$, and $\alpha\sim0.28$ for $4.0\leq z < 5.0$. The typical uncertainty on $\sigma_\mathrm{int}$ is $\sim0.05$, implying that although the intrinsic scatters themselves are statistically significant, the variations of $\lesssim0.02$ (seen in the inset plots) are not. Given the small error bars on the individual $\mu$ and metallicity measurements, the large uncertainties in $\sigma_\mathrm{int}$ likely stem from the small sample sizes in each redshift bin. Even when all three redshift bins are merged to form a larger $1.4\leq z < 5.0$ sample, the variations in $\sigma_\mathrm{int}$ stay at $\sim0.02$, while the uncertainties marginally improve to $\sim0.04$. We therefore caution against drawing strong conclusions from the exact value of the best-fit $\alpha$ or inferring redshift evolution between redshift bins. That being said, we note that these $\alpha$ values are systematically smaller than those derived from stacked analyses ($\alpha\sim0.55–0.66$; e.g., \citealt{Curti2020,Sanders2021,AM2013}), but broadly consistent with the lower values reported in studies based on individual galaxy measurements ($\alpha\sim0.2–0.3$; e.g., \citealt{Mannucci2010} and subsequent work). 

Although a non-zero $\alpha$ reduces the scatter relative to the MZR (or $\alpha$=0), the improvement is modest, at $\sim10\%$ in the lowest two redshift bins and $\sim3\%$ for $4.0\leq z < 5.0$. The degree of this tightening is small compared to the typical improvements of $\sim30-70\%$ in the local Universe \citep{Mannucci2010,AM2013,Curti2020}. In some cases, particularly for $1.4\leq z < 2.7$ and $4.0\leq z < 5.0$, larger values of $\alpha$ (e.g., $\alpha=0.60$) increase the scatter, indicating that a strong SFR dependence is not favored by the data. Overall, the typical intrinsic scatters in our study appear smaller than those evaluated with direct $T_e$-based metallicities (e.g., $\sigma_\mathrm{int}=0.26$ at $z\sim3.2$ in \citealt{Stanton2026}).

We parameterized the best-fit relation in each redshift bin as
\begin{equation}
    12+\log(\mathrm{O/H}) = m \times \mu_{\alpha,9} + Z_{\mu,9},
\end{equation}
where $\mu_{\alpha,9}=\mu_{\alpha}-9$ and $Z_{\mu,9}$ is the \metallicity{} at $\mu_\alpha = 9$. The best-fit parameters can be found in the upper left corners of the panels in Figure~\ref{fig:FMR_alpha_fit}. We find that the normalization of this relation decreases with increasing redshift, similar to the evolution of the MZR. Although combining all 182 galaxies for this analysis would improve the statistical constraints on the scatter, the presence of a redshift evolution in the metallicity normalization would introduce additional scatter across a broad redshift range. This added scatter could obscure the effects of varying $\alpha$, thus motivating our choice to perform the analysis within separate redshift bins. 

The relatively small best-fit $\alpha$ values and the modest reduction in intrinsic scatter compared to the MZR are consistent with the shallow anti-correlations between SFR and metallicity identified in Section~\ref{subsubsec:results-FMR-anticorr}. Together, these findings indicate that while an FMR-like relation is already in place by $z\sim5$, the coupling between star formation activity and chemical enrichment is significantly weaker than that observed locally. 

\section{Discussion} \label{sec:discussion}

\subsection{Comparing our sample to past studies} \label{subsec:disc-compare_sample}

In this work, we analyzed the evolution of the MZR and FMR at high redshift, adding to the growing body of studies enabled by JWST \citep[e.g.,][]{Nakajima2023,Curti2024,Chemerynska2024,Sarkar2025,Stanton2026}. A key advantage of our analysis is the use of stacked spectra, which allows us to derive metallicities without requiring significant detections of fainter emission lines (e.g., H$\beta$, \oii\, and \neiii) to perform dust attenuation corrections and estimate metallicities for individual galaxies. This methodology removes a common source of selection bias toward galaxies with elevated SFRs relative to the SFMS and correspondingly lower metallicities at fixed stellar mass. Our sample is not only more representative of the underlying star-forming galaxy population, as seen from the comparison to the SFMS in Figure~\ref{fig:SFMS}, but also the largest currently available for studies of the MZR and FMR at $z\gtrsim2$.

The large sample size allows us to divide the data into relatively fine redshift and stellar mass bins. Adjacent bins are typically separated by $\sim0.5$~dex in median stellar mass, enabling us to trace the evolution of these relations with greater resolution than many previous studies. While some recent works probe broader redshift and mass ranges, their smaller sample sizes often require combining galaxies across wide intervals in both redshift and stellar mass. For example, the highest-mass bin in \citet{Curti2024} spans nearly the full stellar mass range covered by our $4.0\leq z<5.0$ sample. Additionally, several recent JWST studies have focused primarily on the extreme low-mass regime \citep[e.g.,][]{Curti2024,Chemerynska2024,Chakraborty2025}, whereas our sample probes the more typical stellar mass range occupied by the bulk of star-forming galaxies at these epochs, thereby bridging current low-mass and higher-mass constraints on early galaxy chemical evolution.

\subsection{Comparing our MZR to past studies} \label{subsec:disc-compare_MZR_obs}

\begin{figure}[hb]
    \centering
    \includegraphics[width=0.49\textwidth]{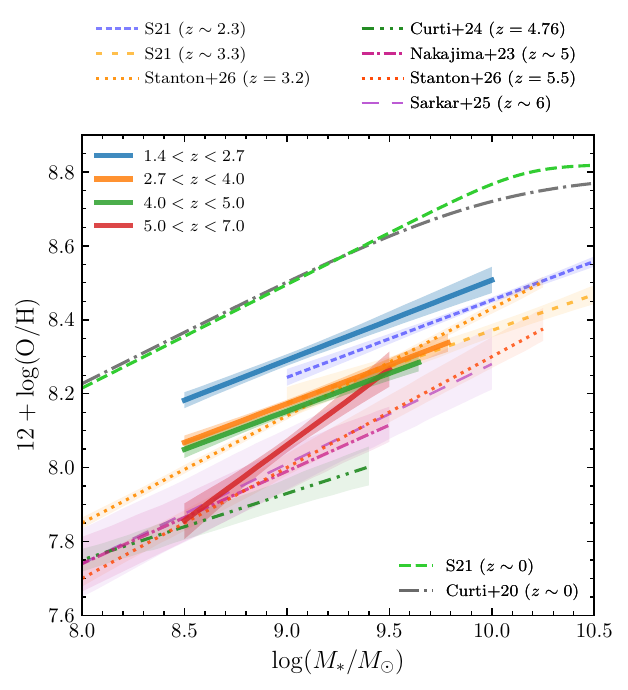}
    \caption{Comparison of the MZRs derived in this work (solid lines), with those reported in literature. Local relations from \citetalias{Sanders2021} and \citet{Curti2020} based on SDSS stacks are denoted in lime green dashed and gray dot-dashed lines. We also display MZR fits across different epochs, including the $z\sim2.3$ (blue closely spaced dashed line) and $z\sim3.3$ (orange widely spaced dashed line) relations from MOSDEF \citepalias{Sanders2021}, the $z=3.2$ light orange dotted line) and $z=5.5$ (orange-red dotted line) from EXCELS \citep{Stanton2026}, the $z=4.8$ relation (dark green double-dot-dashed line) from JADES+CEERS \citep{Curti2024}, the $z\sim5.0$ relation (red-violet dot-dashed line) from ERO+GLASS+CEERS \citep{Nakajima2023}, and the $z\sim6$ relation (purple wide dashed line) from JADES+PRIMAL \citep{Sarkar2025}. Our MZRs at $z\lesssim4$ agree well with those from \citetalias{Sanders2021} and \citet{Stanton2026}, while the normalization is $\sim0.2$~dex higher than that from \citet{Curti2024} at $z\sim4.8$. Our MZR at $z\gtrsim5$ appears steeper than those from literature, but populates the general range of metallicities spanned by other studies at the lower-mass end. 
    \label{fig:MZR_obs}}
\end{figure}

We compare our MZR fits to past studies in Figure~\ref{fig:MZR_obs}. As discussed in Section~\ref{subsec:results-MZR}, we measure an MZR slope of $\gamma \sim 0.21 \pm 0.03$, which holds across $1.4<z<5$, and which falls within the range of values reported in literature. In particular, our results agree closely with studies adopting strong-line metallicity calibrations similar to ours, including \citetalias{Sanders2021} using the updated calibrations from \citetalias{Sanders2025}. Slightly steeper slopes have been reported by \citet{Nakajima2023} ($\gamma = 0.25 \pm 0.03$ for $4<z<10$), \citet{He2024} ($\gamma = 0.26 \pm 0.04$ at $z\sim2.88$), and \citet{Khostovan2025} ($\gamma = 0.27 \pm 0.04$ at $z\sim2.28$), though all consistent with our results within $1.5\sigma$. Other works span a wider range, from the steeper relations of \citet{Stanton2026} ($\gamma = 0.29 \pm 0.01$ at $z\sim3.2$) to the flatter slopes found by \citet{Curti2024} ($\gamma = 0.18 \pm 0.03$ for $3<z<6$). Within a given study \citep[e.g.,][]{Sanders2021,Li2023,Stanton2026}, the slope of the MZR appears to stay roughly the same across these redshifts. 

At $z\gtrsim5$, however, our measured MZR steepens relative to the lower redshift bins to $\gamma = 0.41$, which is in contrast to many studies that find either little evolution in slope or even flattening toward high redshift \citep[e.g.,][]{Nakajima2023,Curti2024}. For instance, \citet{Sarkar2025} report $\gamma \sim$ 0.28, 0.23, and 0.21 at $z\sim$ 5, 7, and 9, respectively. \citet{Kotiwale2026} also find a flat $\gamma = 0.12 \pm 0.08$ for $5<z<7$ based on direct-$T_e$ measurements of metallicities. On the other hand, \citet{Chemerynska2024} measure a comparably steep slope of $\gamma = 0.39 \pm 0.02$ at $6<z<8$. 

Similar to the slope of the MZR at lower redshifts, the normalization of our relations in the lower $1.4<z<4.0$ range are in broad agreement with those in literature, particularly among works using line diagnostics dominated by O3, O32, and Ne3O2. For instance, our recalculated MZR of the MOSDEF stacks from \citetalias{Sanders2021}, as reported in Table~\ref{tab:MZR_fits}, have $Z_9=8.26\pm0.02$ and $Z_9=8.18\pm0.04$ at $z\sim2.08$ and $z\sim3.24$ after accounting for redshift evolution. \citet{Stanton2026} report $Z_9=8.14\pm0.01$ at $z\sim3.2$. These measurements are consistent within $2\sigma$ uncertainty of our results at $1.4<z<4.0$. On the other hand, studies that mainly employ the direct $T_e$-based method to measure metallicities find lower normalizations at fixed \mast\ (e.g. $Z_9=8.17\pm0.06$ at $z\sim2.28$ from \citealt{Khostovan2025}). This discrepancy may arise because faint \oiii\ auroral emission lines required for $T_e$-based metallicities are more readily detected in low-metallicity galaxies, where higher electron temperatures enhance line emissivities.

In the $4.0<z<5.0$ range, our MZR normalization does not decrease much from the $2.7<z<4.0$ relation. This normalization is over 0.2 dex higher compared to the $z=4.76$ relation of \citet{Curti2024} with $Z_9=7.93$ (inferred from their reported values of $Z_8=7.75$ and $\gamma=0.18$). These discrepancies may arise from differences in sample selection. For example, \citet{Curti2024} probe lower stellar masses (down to $\sim 10^7$~\msol{}) and target galaxies that lie systematically above the SFMS at fixed mass by about 0.5~dex. If enhanced SFRs correspond to lower metallicities at fixed mass, as we demonstrate in Section~\ref{subsubsec:results-FMR-anticorr}, such selection effects could contribute to lower observed metallicity normalizations. 
However, even if we shift the \citet{Curti2024} MZR upwards by 0.1 dex in metallicity to account for this anti-correlation at $4 < z < 5$, we cannot fully erase this 0.2 dex discrepancy.

At $z \gtrsim 5$, many past studies appear to agree on the metallicity normalizations, all converging at $12+\log(\mathrm{O/H})\sim 7.8$ at $10^{8.5}$~\msol{} \citep{Nakajima2023,Sarkar2025,Stanton2026}. Due to the lower stellar masses and thus lower metallicity ranges probed by these studies, some of them adopt different sets of strong-line calibrations. For example, \citet{Nakajima2023} and \citet{Sarkar2025} primarily use R23 and/or O3 in the lower-branch regime, where O32 and Ne3O2 instead plateau. 
Interestingly, the low-mass end of our $5.0<z<7.0$ MZR is broadly consistent with these previous measurements, whereas the high-mass end aligns more closely with the $2.7<z<4.0$ and $4.0<z<5.0$ relations. 
These features of the $5.0<z<7.0$ MZR reflect a combination of physical effects: rapid chemical maturation of higher-mass galaxies prior to $z\sim5$, bringing them closer to their lower redshift counterparts in metallicity, while lower-mass systems at these early epochs are still far from equilibrium due to strong gas accretion and outflows that hinder the buildup of metals.
However, because the $5.0<z<7.0$ relation is constrained by only two stacked points, the current data do not yet allow us to concretely verify this possible mass-dependent chemical enrichment history.

Regardless of the behavior at the highest redshifts, we find clear evidence that the MZR slope evolves only weakly from $z\sim4.4$, while the normalization exhibits smooth evolution starting at $z\sim3.24$. The inferred normalization evolution, $d\log(\mathrm{O/H})/dz = -0.10\pm0.02$, is highly consistent with previous measurements from \citetalias{Sanders2021} and \citet{Li2023}, who find comparable rates of metallicity decline with redshift. Beyond $z\gtrsim3.24$, however, we do not observe significant further decreases in the overall normalization, in agreement with \citet{Nakajima2023}, who similarly report little evolution over $4<z<10$. This mild evolution contrasts with studies such as \citet{Sarkar2025} and \citet{Stanton2026}, which infer continued metallicity declines toward higher redshift, with rates of $d\log(\mathrm{O/H})/dz = -0.06\pm0.01$ and a total decrease of $\sim0.14$~dex from $z\sim3.2$ to $z\sim5.5$, respectively.

\subsection{Comparing our MZR to simulations} \label{subsec:disc-compare_MZR_sim}

\begin{figure}[hb]
    \centering
    \includegraphics[width=0.49\textwidth]{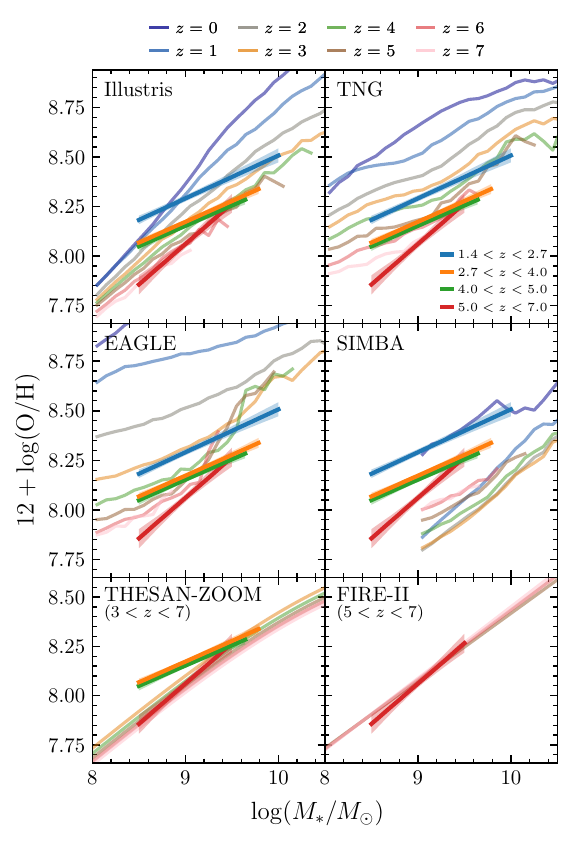}
    \caption{Comparison of the MZRs derived in this work (solid lines), with simulations. Each panel shows the median MZR for resolved galaxies at integer increments of redshift at $z=0-7$ in Illustris (top left), TNG (top right), EAGLE (middle left), and SIMBA (middle right) curated in \citet{Garcia2025}, $z=3-7$ in THESAN-ZOOM \citep[bottom left;][]{McClymont2026}, and $z=5-7$ in FIRE-II \citep[bottom right;][]{Marszewski2024}.
    \label{fig:MZR_sim}}
\end{figure}

In Figure~\ref{fig:MZR_sim}, we compare our observed MZR measurements with predictions from several cosmological simulations, including Illustris \citep{Vogelsberger2013}, IllustrisTNG\citep{Nelson2019}, EAGLE \citep{Schaye2015}, and SIMBA \citep{Dave2019} compiled by \citet{Garcia2025}, evaluated at integer redshift intervals at $z=0-7$. We additionally include the high-redshift MZR relations from THESAN-ZOOM and FIRE-II from \citet{McClymont2026} and \citet{Marszewski2024}, shown over the redshift ranges $3<z<7$ and $5<z<7$, respectively. Overall, our measurements fall within the general envelope spanned by the simulations, although no single model simultaneously reproduces both the observed normalization and slope across all redshifts.

At $z<5$, different simulations capture different aspects of our observed MZR. Illustris provides the closest agreement in normalization across different redshifts, particularly around $M_\ast \sim 10^{9-9.5}$~\msol{}, but predicts substantially steeper slopes than observed. In contrast, IllustrisTNG reproduces the observed slopes more accurately while systematically overpredicting metallicities at fixed stellar mass. EAGLE and SIMBA generally fall between these extremes, though neither fully match both the normalization and shape of the observed relation. Interestingly, the $z>5$ MZRs in both TNG and EAGLE exhibit a flat relation at low masses followed by a steeper rise above $M_\ast \sim 10^{9.25}$~\msol{}. Indeed, if we average the flat and steep slopes of the EAGLE $z\sim6$ MZR, the resultant averaged slope would qualitatively resemble that suggested by our $5<z<7$ measurements.

The simulations also differ in their predicted redshift evolution in normalization. Most models show relatively smooth decreases in metallicity normalization toward high redshift, while EAGLE predicts a partial slowing of this evolution at early times. At $z>5$, our measurements are in comparatively good agreement with the simulated relations, particularly those from THESAN-ZOOM and FIRE-II. This agreement may indicate that high-resolution zoom-in simulations, which more explicitly implement bursty star formation, metal transport, and feedback processes in low-mass galaxies, are better able to reproduce the physical conditions governing early chemical enrichment. 

The MZRs derived from simulations are sensitive to their implementation of baryonic physics, including feedback efficiencies, gas inflow and outflow prescriptions, metal mixing, and star formation histories. The lack of consensus between the predictions of the MZR and the mismatch with observational data across redshifts suggest that our current models still struggle to fine tune the extent of metal retention and the regulation of star formation across the full galaxy population. Larger observational samples spanning a wider stellar mass range will be necessary to discriminate between competing models and to provide additional constraints for models of galaxy formation.

\subsection{A weak FMR at high redshifts} \label{subsec:disc-FMR}

In the local universe, galaxies are thought to exist in a quasi-equilibrium state in which stellar feedback, as well as gas inflows and outflows, collectively regulate the relationship between stellar mass, metallicity, and SFR, as described by the FMR \citep[e.g.,][]{Mannucci2010,AM2013,Curti2020}. As a logical extension, considerable effort has been devoted to determining whether this regulatory relationship remains invariant across cosmic time, and if so, when it becomes established. In the era of JWST surveys, an emerging picture has begun to develop where high-redshift galaxies earlier than $z\sim4$ are systematically offset toward lower metallicities at fixed stellar mass and SFR compared to the local FMR \citep[e.g.,][]{Nakajima2023,Curti2024}. Our results support and strengthen this growing evidence for an evolving FMR across cosmic time.

One of the key findings of this paper is the presence of a statistically significant anti-correlation between SFR residuals relative to the SFMS and metallicity residuals relative to the MZR within individual redshift bins spanning $z\sim2–5$. Previously, using high signal-to-noise stacked spectra from MOSDEF, \citetalias{Sanders2021} (and \citealt{Sanders2018}) reported evidence for such an anti-correlation at $z\sim2.3$. In contrast, studies based on individual galaxies, such as \citet{Korhonen2025} at $z\sim2.3$ and \citet{Stanton2026} out to $z\sim5.5$, do not recover statistically significant anti-correlations. Our analysis therefore provides the first evidence from high signal-to-noise stacked spectra that this relationship persists beyond $z\sim2.3$. The existence of this anti-correlation demonstrates that a connection between star formation activity and chemical enrichment remains established within the galaxy population at each epoch. The trend is present across both low- and high-mass systems and is especially pronounced in the lower-redshift bins. Nevertheless, the slope of the relation is systematically shallower than observed locally, indicating that metallicity depends less strongly on SFR at fixed stellar mass in the early universe. Notably, while we do not recover this anti-correlated relation in the full $5.0\leq z<7.0$ bin, the two high-mass stacks display a negative trend similar to those of the lower-redshift bins, though consistent with zero slope within the $1\sigma$ uncertainties. This marginal evidence adds to our discussion in Section~\ref{subsec:disc-compare_MZR_obs} of the mismatch in the behavior of the MZR at high- and low-masses in the $5.0\leq z<7.0$ range. The galaxy population at this redshift appears to exist in two different evolutionary stages --- one where the physical processes known to regulate galaxy growth at lower redshifts begin to be established in galaxies with higher stellar masses, and another at lower masses where turbulent processes still dominate galaxy assembly as seen at even higher redshifts. 

Another piece of evidence for an evolving \mast{}-Z-SFR relation comes from our $\mu_\alpha$-based analysis of the FMR. We find that the value of $\alpha$ that minimizes the intrinsic scatter in the MZR is consistently lower than the values derived from stacked SDSS galaxies of $\alpha\sim0.55-0.66$ \citep[e.g.,][]{AM2013,Curti2020,Sanders2021}, though broadly consistent with the original \citet{Mannucci2010} result of $\alpha=0.32$, as well as $\alpha=0.33$ at high redshift from \citet{Stanton2026}. Moreover, incorporating SFR produces only a marginal reduction in scatter ($\sim3-10\%$) compared to the substantial tightening observed at $z\sim0$ ($\gtrsim50\%$; \citealt{Mannucci2010,AM2013,Curti2020}). These findings are also qualitatively consistent with recent cosmological simulations, most of which find that $\alpha$ evolves with redshift and that the coupling between metallicity and SFR weakens toward earlier cosmic times \citep[e.g.,][]{Garcia2025}. 
However, some recent studies report little to no evidence for an FMR at high redshift and low stellar masses \citep[e.g.,][]{Korhonen2025}, and even indications of a trend opposite to the conventional formulation \citep{Laseter2025}. 
Meanwhile, \citetalias{Sanders2021} find a factor of 1.3--2 decrease in scatter at $z\sim2.3$ and $3.3$, similar to the improvement seen at $z\sim0$. While this result appears in contrast with ours, we note that the MOSDEF sample probes systematically higher stellar masses than those considered in this work. In a recent study, \citet{Laseter2025} suggest that the FMR weakens towards lower stellar masses and may even reverse at the lowest masses, even at $z\sim0$. Since our sample extends to lower stellar masses, our weaker evidence for an FMR may reflect mass dependence in the \mast{}-Z-SFR relation. This interpretation is consistent with a scenario in which low-mass galaxies at high redshift have not yet settled into the tight equilibrium relation observed in more massive systems. 

A possible explanation for the weakened coupling between stellar mass, metallicity, and SFR at high redshift is that the physical processes governing star formation and chemical enrichment become increasingly decoupled. Using the IllustrisTNG simulation suite, \citet{Torrey2018} showed that the local FMR emerges when offsets from the SFMS and MZR vary coherently on similar timescales. However, high-redshift galaxies are observed to experience more bursty, feedback-driven star formation histories, causing SFRs to fluctuate on relatively short timescales \citep[e.g.,][]{Faucher2018,Clarke2025,Cole2025,Perry2025}. In contrast, the gas-phase metallicity may evolve more gradually, on timescales comparable to the dynamical time of the dark matter halo, particularly if outflows remove gas with metallicities similar to that of the ISM \citep{Torrey2018}. This mismatch in variability timescales would naturally weaken the observed coupling between SFR and metallicity at fixed stellar mass. Additionally, the substantially larger gas fractions present in high-redshift galaxies may buffer or dilute the metallicity response to pristine gas inflows during periods of elevated star formation \citep{Somerville2015,Dave2017}. In this picture, stochastic star formation and strong outflows may represent the typical conditions in galaxies rather than exceptional transient states. Such a scenario is qualitatively consistent with both observations and simulations indicating that high-redshift galaxies possess large gas reservoirs, elevated SFRs, and strong feedback-driven outflows \citep[e.g.,][]{Hayward2017,Sparre2017,Tacconi2020,Tacchella2020,Simmonds2025}.

\subsection{The required parameters for future MZR/FMR studies at high redshift} \label{subsec:disc-future_sample}

Altogether, recent works reveal an increasingly diverse picture of chemical enrichment and galaxy evolution, with different studies probing distinct regions of stellar mass, star formation activity, and cosmic time. Current observational studies of galaxy chemical evolution, enabled by JWST, analyze samples that span a large dynamic range in stellar mass, reaching even as low as $10^6\ M_\odot$ and as early as $z=10$ \citep[e.g.,][]{Chemerynska2024,Stanton2026,Pollock2026}. However, the samples analyzed comprise a relatively small number of galaxies in comparison with ground-based observational surveys such as FMOS-COSMOS, MOSDEF, and KBSS-MOSFIRE, for which the large sample size enables a fine sampling of galaxy properties in both stellar mass and redshift \citep{Zahid2014,Sanders2021,Topping2021,Korhonen2025}. The comparison of the current literature with works at cosmic noon highlights the need for a large representative sample star-forming galaxies with metallicity estimates across cosmic time. 

Our results suggest that our sample sizes at $z\sim1.4$ to $z\sim5.0$ can be sufficient to constrain low-order relations such as the MZR. However, higher-order analyses, such as deviations from the SFMS and MZR, as well as measurements of the intrinsic scatter of the FMR, require substantially larger samples. Considering the MZR, for example, matching the signal-to-noise of each stack to those of \citetalias{Sanders2021} by requiring at least 50 galaxies per stack, while also constructing three stacks per redshift range to robustly constrain the MZR slope, would require $\sim150$ galaxies per redshift range. Alternatively, reaching a fine stellar mass resolution of $\delta\log(M_\ast/M_\odot)\sim0.3$ would require approximately five stacks per redshift bin; even with a more modest 30 galaxies per bin, this would similarly demand a minimum of $\sim150$ galaxies per redshift range. Achieving such sample sizes, even across a mass range of only $10^{8.5-10}\,M_\odot$, would therefore require roughly a factor of 2–3 increase in the available data. Expanding to a broader dynamic range in stellar mass would further increase the required sample size. Ultimately, a sample size at least as extensive as existing ground-based surveys at cosmic noon is needed. An extended sample would provide sufficient mass and temporal resolution to robustly trace galaxy evolution through cosmic time, as well as constrain simulations and models of galaxy growth.

\section{Conclusion} \label{sec:conclusion}

In this study, we have used the JADES DR3 dataset to analyze the star-forming galaxy MZR and FMR in the redshift range $1.4 < z< 7$ for galaxies at $M_*>10^{8.5}\ M_\odot$. We implement the novel strong-line calibrations from \citetalias{Sanders2025} and conduct a robust stacking analysis to 601 (418) galaxies across four bins in redshift and 14 (10) bins in stellar mass. This work leverages the largest sample at $z\gtrsim 2$ to date, providing insights into the chemical evolution of the high-redshift galaxy population. We summarize our main conclusions as follows:

\begin{enumerate}
    \item We fit the MZR in the redshift ranges $1.4\leq z < 2.7$, $2.7\leq z < 4.0$, $4.0\leq z < 5.0$, and $5.0\leq z < 7.0$ (see Table~\ref{tab:MZR_fits}). The slope of the MZR stays constant between $1.4\leq z < 5.0$ at $\gamma\sim0.21$ and steepens to $\gamma=0.41$ in the $5.0\leq z < 7.0$ range. The normalization decreases from $Z_9\sim8.5$ at $z\sim0$ to 8.29 and 8.17 at $z\sim2.08$ and $z\sim3.24$, corresponding to a smooth evolution of $d\log(\mathrm{O/H})/dz = -0.10\pm0.02$. The decrease in normalization from $z\sim3.24$ to $z\sim4.40$ ($Z_9=8.15$) of 0.02 dex is very mild. While we obtain $Z_9=8.06$ for the $5.0\leq z < 7.0$ relation, the high-mass end matches up with the MZRs at $z\sim3.24$ and $z\sim4.40$, suggesting that the low-mass end drives the steepening of the MZR in this highest redshift range.
    \item We find that the $z\gtrsim2$ FMR is offset below the $z\sim 0$ FMR by $\sim -0.15$ dex, suggesting a redshift-dependent evolution of the FMR. This finding is in contrast with previous works that suggest a non-evolving FMR with redshift out to $z\sim3$ \citepalias{Sanders2021}, but is in agreement with recent works suggesting that galaxies may exhibit lower metallicities at fixed SFR and $M_*$ than would be suggested by the $z\sim 0 $ FMR \citep[e.g.,][]{Nakajima2023,Curti2024}.
    \item We recover statistically significant anti-correlations between SFMS offsets and MZR offsets in the $1.4\leq z < 2.7$, $2.7\leq z < 4.0$, and $4.0\leq z < 5.0$ redshift bins, detecting this anti-correlation for the first time at $z\gtrsim3$. The slopes of $\nu=-0.13\pm0.05$, $\nu=-0.08\pm0.04$, and $\nu=-0.09\pm0.06$ are shallower than that in the local Universe ($\nu\sim-0.27$). A relationship between SFR and metallicity appears to be established by these epochs, albeit with a weaker coupling than observed at $z\sim0$.
    \item We estimate the FMR projection parameter to be $\alpha=0.26-0.38$, falling below the more typically assumed value of $\alpha\sim0.6$ of the local Universe. Accordingly, the decrease in the MZR scatter that occurs when incorporating galaxy SFRs (i.e., constructing the FMR) is marginal, varying between 3-11\%, as opposed to the up to 70\% decrease in scatter in the local Universe \citep[e.g.,][]{AM2013,Curti2020}. This lower value of $\alpha$ suggests a weaker anti-correlation between SFR and metallicity at fixed stellar mass than is typically assumed. However, this lower $\alpha$ value agrees well with studies that estimate this parameter from individual, rather than stacked, galaxy measurements.
\end{enumerate}

\section{Data Availability}
The High-Level Science Products (HLSP) from FRESCO \citep{2023MNRAS.525.2864O}, JEMS \citep{2023ApJS..268...64W}, and JADES \citep{2023ApJS..269...16R,2023arXiv230602465E} can be accessed via the Mikulski Archive for Space Telescopes (MAST) at the Space Telescope Science Institute (STScI) \citet{https://doi.org/10.17909/gdyc-7g80,https://doi.org/10.17909/fsc4-dt61,https://doi.org/10.17909/8tdj-8n28}.

\begin{acknowledgments}
We would like to acknowledge the JADES team for their efforts in designing, executing, and making public their JWST/NIRSpec and JWST/NIRCam survey data. We also acknowledge support from NASA grants JWST-GO-01914 and JWST-GO-03833, and NSF AAG grants 2009313, 2009085, 2307622, and 2307623. N. L. is supported by the National Science Foundation Graduate Research Fellowship Program. 
This work is based on observations made with the NASA/ESA/CSA James Webb Space Telescope as well as the NASA/ESA Hubble Space Telescope. The data were obtained from the Mikulski Archive for Space Telescopes at the Space Telescope Science Institute (STScI), which is operated by the Association of Universities for Research in Astronomy, Inc., under NASA contract NAS5-03127 for JWST and NAS 5–26555 for HST. Data were also obtained from the DAWN JWST Archive maintained by the Cosmic Dawn Center. The specific observations analyzed can be
accessed via \dataset[doi:10.17909/8tdj-8n28]{https://archive.stsci.edu/doi/resolve/resolve.html?doi=10.17909/8tdj-8n28}, \dataset[doi:10.17909/gdyc-7g80]{https://archive.stsci.edu/doi/resolve/resolve.html?doi=10.17909/gdyc-7g80}, and \dataset[doi:10.17909/fsc4-dt61]{https://archive.stsci.edu/doi/resolve/resolve.html?doi=10.17909/fsc4-dt61}. This work used computational and storage services associated with the Hoffman2 Cluster which is operated by the UCLA Office of Advanced Research Computing’s Research Technology Group.
\end{acknowledgments}

\bibliography{main}{}
\bibliographystyle{aasjournal}


\end{document}

%% file: table_mzr_stacks_nosfr.tex
\begin{deluxetable*}{cccccccc}[h!t]
\tablecaption{Properties of stacked spectra for the MZR samples in bins of M$_{\ast}$ \label{tab:MZR_stacks}}

\tablehead{
    $\log\left(\frac{\mathrm{M}_{\ast}}{\mathrm{M}_{\odot}}\right)$\tablenotemark{a} &
    $N_\mathrm{gal}$\tablenotemark{b} &
    $\log(\mathrm{O3})$ &
    $\log(\mathrm{O32})$ &
    $\log(\mathrm{Ne3O2})$ &
    $\log(\mathrm{R23})$ &
    $\log(\mathrm{N2})$ &
    $12+\log(\mathrm{O/H})$
}

\startdata
\hline\hline
\multicolumn{8}{c}{$1.4 \leq z < 2.7$ in bins of M$_{\ast}$ } \\
\hline
$^{\ast}$$8.18_{-0.03}^{+0.04}$ & $47$ & $0.72_{-0.04}^{+0.04}$ & $0.54_{-0.07}^{+0.07}$ & $-0.40_{-0.08}^{+0.08}$ & $0.92_{-0.04}^{+0.04}$ & $-1.84_{-0.01}^{+0.01}$ & $8.11_{-0.04}^{+0.04}$ \\
$8.74_{-0.03}^{+0.03}$ & $51$ & $0.67_{-0.03}^{+0.03}$ & $0.41_{-0.05}^{+0.04}$ & $-0.64_{-0.10}^{+0.09}$ & $0.91_{-0.04}^{+0.04}$ & $-1.51_{-0.11}^{+0.09}$ & $8.23_{-0.03}^{+0.03}$ \\
$9.12_{-0.02}^{+0.02}$ & $51$ & $0.63_{-0.03}^{+0.03}$ & $0.22_{-0.04}^{+0.04}$ & $-0.83_{-0.07}^{+0.06}$ & $0.92_{-0.04}^{+0.04}$ & $-1.24_{-0.04}^{+0.04}$ & $8.32_{-0.02}^{+0.02}$ \\
$9.68_{-0.02}^{+0.02}$ & $50$ & $0.49_{-0.03}^{+0.03}$ & $-0.03_{-0.04}^{+0.04}$ & $-1.11_{-0.09}^{+0.07}$ & $0.86_{-0.03}^{+0.03}$ & $-0.78_{-0.02}^{+0.02}$ & $8.44_{-0.01}^{+0.01}$ \\
\hline\hline
\multicolumn{8}{c}{$2.7 \leq z < 4.0$ in bins of M$_{\ast}$ } \\
\hline
$^{\ast}$$8.17_{-0.04}^{+0.04}$ & $51$ & $0.73_{-0.02}^{+0.02}$ & $0.69_{-0.04}^{+0.04}$ & $-0.48_{-0.08}^{+0.07}$ & $0.92_{-0.02}^{+0.02}$ & $-1.79_{-0.13}^{+0.10}$ & $8.09_{-0.03}^{+0.03}$ \\
$8.71_{-0.03}^{+0.03}$ & $44$ & $0.75_{-0.02}^{+0.02}$ & $0.61_{-0.03}^{+0.03}$ & $-0.46_{-0.06}^{+0.05}$ & $0.94_{-0.02}^{+0.02}$ & $-1.57_{-0.08}^{+0.06}$ & $8.11_{-0.02}^{+0.02}$ \\
$9.04_{-0.02}^{+0.02}$ & $44$ & $0.73_{-0.02}^{+0.02}$ & $0.47_{-0.03}^{+0.03}$ & $-0.58_{-0.05}^{+0.05}$ & $0.95_{-0.02}^{+0.02}$ & $-1.33_{-0.04}^{+0.04}$ & $8.18_{-0.02}^{+0.02}$ \\
$9.57_{-0.03}^{+0.03}$ & $43$ & $0.67_{-0.01}^{+0.01}$ & $0.28_{-0.02}^{+0.02}$ & $-0.84_{-0.04}^{+0.04}$ & $0.93_{-0.01}^{+0.01}$ & $-0.95_{-0.02}^{+0.02}$ & $8.29_{-0.01}^{+0.01}$ \\
\hline\hline
\multicolumn{8}{c}{$4.0 \leq z < 5.0$ in bins of M$_{\ast}$ } \\
\hline
$^{\ast}$$8.14_{-0.04}^{+0.07}$ & $32$ & $0.70_{-0.02}^{+0.02}$ & $0.69_{-0.04}^{+0.03}$ & $-0.35_{-0.05}^{+0.04}$ & $0.89_{-0.02}^{+0.02}$ & $-1.57_{-0.08}^{+0.06}$ & $8.06_{-0.03}^{+0.03}$ \\
$8.73_{-0.03}^{+0.03}$ & $37$ & $0.76_{-0.02}^{+0.02}$ & $0.62_{-0.03}^{+0.03}$ & $-0.45_{-0.03}^{+0.03}$ & $0.95_{-0.02}^{+0.02}$ & $-1.51_{-0.06}^{+0.05}$ & $8.10_{-0.02}^{+0.01}$ \\
$9.35_{-0.06}^{+0.06}$ & $36$ & $0.72_{-0.02}^{+0.01}$ & $0.37_{-0.02}^{+0.02}$ & $-0.66_{-0.03}^{+0.03}$ & $0.96_{-0.02}^{+0.02}$ & $-1.08_{-0.02}^{+0.02}$ & $8.22_{-0.01}^{+0.01}$ \\
\hline\hline
\multicolumn{8}{c}{$5.0 \leq z < 7.0$ in bins of M$_{\ast}$ } \\
\hline
$^{\ast}$$8.20_{-0.04}^{+0.04}$ & $53$ & $0.73_{-0.01}^{+0.01}$ & $0.92_{-0.03}^{+0.03}$ & $-0.11_{-0.04}^{+0.04}$ & $0.89_{-0.01}^{+0.01}$ & $-1.78_{-0.13}^{+0.10}$ & $7.69_{-0.04}^{+0.05}$ \\
$8.71_{-0.04}^{+0.04}$ & $31$ & $0.76_{-0.02}^{+0.02}$ & $0.77_{-0.03}^{+0.04}$ & $-0.24_{-0.05}^{+0.04}$ & $0.93_{-0.02}^{+0.02}$ & $-1.46_{-0.07}^{+0.06}$ & $7.94_{-0.04}^{+0.03}$ \\
$9.21_{-0.05}^{+0.04}$ & $31$ & $0.76_{-0.01}^{+0.01}$ & $0.51_{-0.02}^{+0.02}$ & $-0.53_{-0.03}^{+0.03}$ & $0.97_{-0.01}^{+0.01}$ & $-1.21_{-0.03}^{+0.03}$ & $8.15_{-0.01}^{+0.01}$ \\
\hline\hline
\enddata
\tablenotetext{a}{Median stellar mass of galaxies in each bin.}
\tablenotetext{b}{Number of galaxies in each bin.}
\end{deluxetable*}

%% file: table_fmr_stacks.tex
\begin{deluxetable*}{ccccccccc}[h!t]
\tablecaption{Properties of stacked spectra for the FMR samples in bins of M$_{\ast}$ \label{tab:FMR_stacks}}

\tablehead{
    $\log\left(\frac{\mathrm{M}_{\ast}}{\mathrm{M}_{\odot}}\right)$\tablenotemark{a} &
    $N_\mathrm{gal}$\tablenotemark{b} &
    $\log\left(\frac{\mathrm{SFR}}{\mathrm{M}_{\odot}\,\mathrm{yr}^{-1}}\right)$\tablenotemark{c} &
    $\log(\mathrm{O3})$ &
    $\log(\mathrm{O32})$ &
    $\log(\mathrm{Ne3O2})$ &
    $\log(\mathrm{R23})$ &
    $\log(\mathrm{N2})$ &
    $12+\log(\mathrm{O/H})$
}

\startdata
\hline\hline
\multicolumn{9}{c}{$1.4 \leq z < 2.7$ in bins of M$_{\ast}$ } \\
\hline
$^{\ast}$$8.19_{-0.06}^{+0.08}$ & $21$ & $-0.08_{-0.06}^{+0.06}$ & $0.74_{-0.04}^{+0.04}$ & $0.77_{-0.06}^{+0.07}$ & $-0.26_{-0.09}^{+0.08}$ & $0.91_{-0.03}^{+0.04}$ & $-1.94_{-0.01}^{+0.01}$ & $7.96_{-0.08}^{+0.06}$ \\
$8.79_{-0.04}^{+0.05}$ & $32$ & $-0.03_{-0.06}^{+0.04}$ & $0.73_{-0.03}^{+0.03}$ & $0.44_{-0.05}^{+0.05}$ & $-0.64_{-0.07}^{+0.06}$ & $0.95_{-0.03}^{+0.03}$ & $-1.49_{-0.07}^{+0.06}$ & $8.21_{-0.02}^{+0.02}$ \\
$9.24_{-0.02}^{+0.02}$ & $32$ & $0.26_{-0.05}^{+0.05}$ & $0.63_{-0.03}^{+0.03}$ & $0.20_{-0.05}^{+0.04}$ & $-0.82_{-0.06}^{+0.05}$ & $0.92_{-0.04}^{+0.04}$ & $-1.18_{-0.03}^{+0.03}$ & $8.32_{-0.02}^{+0.02}$ \\
$9.80_{-0.02}^{+0.03}$ & $31$ & $0.83_{-0.04}^{+0.03}$ & $0.40_{-0.03}^{+0.03}$ & $-0.15_{-0.04}^{+0.03}$ & $-1.00_{-0.01}^{+0.01}$ & $0.82_{-0.03}^{+0.04}$ & $-0.72_{-0.01}^{+0.01}$ & $8.51_{-0.01}^{+0.01}$ \\
\hline\hline
\multicolumn{9}{c}{$2.7 \leq z < 4.0$ in bins of M$_{\ast}$ } \\
\hline
$^{\ast}$$8.18_{-0.05}^{+0.04}$ & $39$ & $0.11_{-0.04}^{+0.04}$ & $0.73_{-0.02}^{+0.02}$ & $0.74_{-0.03}^{+0.04}$ & $-0.46_{-0.06}^{+0.07}$ & $0.91_{-0.02}^{+0.02}$ & $-1.69_{-0.09}^{+0.08}$ & $8.08_{-0.03}^{+0.03}$ \\
$8.73_{-0.04}^{+0.03}$ & $39$ & $0.15_{-0.05}^{+0.07}$ & $0.76_{-0.02}^{+0.02}$ & $0.61_{-0.03}^{+0.03}$ & $-0.46_{-0.05}^{+0.05}$ & $0.96_{-0.02}^{+0.02}$ & $-1.58_{-0.08}^{+0.07}$ & $8.10_{-0.02}^{+0.02}$ \\
$9.05_{-0.02}^{+0.02}$ & $38$ & $0.33_{-0.07}^{+0.07}$ & $0.74_{-0.02}^{+0.02}$ & $0.50_{-0.03}^{+0.03}$ & $-0.57_{-0.05}^{+0.05}$ & $0.95_{-0.02}^{+0.02}$ & $-1.32_{-0.04}^{+0.03}$ & $8.17_{-0.02}^{+0.02}$ \\
$9.66_{-0.04}^{+0.02}$ & $38$ & $0.82_{-0.02}^{+0.04}$ & $0.65_{-0.01}^{+0.01}$ & $0.26_{-0.02}^{+0.02}$ & $-0.86_{-0.05}^{+0.04}$ & $0.92_{-0.01}^{+0.01}$ & $-0.94_{-0.02}^{+0.02}$ & $8.31_{-0.01}^{+0.01}$ \\
\hline\hline
\multicolumn{9}{c}{$4.0 \leq z < 5.0$ in bins of M$_{\ast}$ } \\
\hline
$^{\ast}$$8.18_{-0.06}^{+0.05}$ & $29$ & $0.26_{-0.03}^{+0.03}$ & $0.74_{-0.02}^{+0.02}$ & $0.71_{-0.03}^{+0.03}$ & $-0.35_{-0.04}^{+0.04}$ & $0.92_{-0.02}^{+0.02}$ & $-1.58_{-0.07}^{+0.06}$ & $8.03_{-0.03}^{+0.02}$ \\
$8.73_{-0.03}^{+0.04}$ & $34$ & $0.50_{-0.06}^{+0.05}$ & $0.77_{-0.01}^{+0.02}$ & $0.65_{-0.03}^{+0.03}$ & $-0.41_{-0.03}^{+0.03}$ & $0.96_{-0.01}^{+0.02}$ & $-1.51_{-0.04}^{+0.04}$ & $8.07_{-0.02}^{+0.02}$ \\
$9.39_{-0.04}^{+0.04}$ & $34$ & $0.91_{-0.03}^{+0.03}$ & $0.72_{-0.01}^{+0.02}$ & $0.37_{-0.02}^{+0.02}$ & $-0.67_{-0.04}^{+0.03}$ & $0.96_{-0.02}^{+0.02}$ & $-1.06_{-0.02}^{+0.02}$ & $8.23_{-0.01}^{+0.01}$ \\
\hline\hline
\multicolumn{9}{c}{$5.0 \leq z < 7.0$ in bins of M$_{\ast}$ } \\
\hline
$^{\ast}$$8.20_{-0.05}^{+0.04}$ & $47$ & $0.47_{-0.03}^{+0.04}$ & $0.73_{-0.01}^{+0.01}$ & $0.97_{-0.03}^{+0.03}$ & $-0.10_{-0.04}^{+0.04}$ & $0.89_{-0.01}^{+0.01}$ & $-1.73_{-0.11}^{+0.10}$ & $7.66_{-0.04}^{+0.04}$ \\
$8.71_{-0.03}^{+0.04}$ & $28$ & $0.62_{-0.07}^{+0.06}$ & $0.75_{-0.01}^{+0.01}$ & $0.78_{-0.03}^{+0.04}$ & $-0.24_{-0.04}^{+0.04}$ & $0.93_{-0.01}^{+0.01}$ & $-1.45_{-0.08}^{+0.07}$ & $7.93_{-0.04}^{+0.04}$ \\
$9.21_{-0.04}^{+0.05}$ & $27$ & $0.97_{-0.04}^{+0.04}$ & $0.76_{-0.01}^{+0.01}$ & $0.51_{-0.02}^{+0.02}$ & $-0.53_{-0.03}^{+0.03}$ & $0.97_{-0.01}^{+0.01}$ & $-1.23_{-0.03}^{+0.03}$ & $8.15_{-0.01}^{+0.01}$ \\
\hline\hline
\enddata
\tablenotetext{a}{Median stellar mass of galaxies in each bin.}
\tablenotetext{b}{Number of galaxies in each bin.}
\tablenotetext{c}{SFR determined from median SFR of constituent galaxies.}
\end{deluxetable*}

%% file: table_anticorr_stacks.tex
\begin{deluxetable*}{ccccccccc}[h!t]
\tablecaption{Properties of stacked spectra for the FMR samples in bins of M$_{\ast}$-$\Delta$SFR \label{tab:anticorr_stacks}}

\tablehead{
    \colhead{$\log\left(\frac{\mathrm{M}_{\ast}}{\mathrm{M}_{\odot}}\right)$\tablenotemark{a}} &
    \colhead{$N_\mathrm{gal}$\tablenotemark{b}} &
    \colhead{$\log\left(\frac{\mathrm{SFR}}{\mathrm{M}_{\odot}\,\mathrm{yr}^{-1}}\right)$\tablenotemark{c}} &
    \colhead{$\log(\mathrm{O3})$} &
    \colhead{$\log(\mathrm{O32})$} &
    \colhead{$\log(\mathrm{Ne3O2})$} &
    \colhead{$\log(\mathrm{R23})$} &
    \colhead{$\log(\mathrm{N2})$} &
    \colhead{$12+\log(\mathrm{O/H})$}
}

\startdata
\hline\hline
\multicolumn{9}{c}{$1.4 \leq z < 2.7$ in bins of M$_{\ast}$-$\Delta$sSFR} \\
\hline
$8.96_{-0.05}^{+0.04}$ & $27$ & $-0.19_{-0.04}^{+0.05}$ & $0.62_{-0.03}^{+0.03}$ & $0.33_{-0.03}^{+0.04}$ & $-0.70_{-0.03}^{+0.03}$ & $0.88_{-0.03}^{+0.03}$ & $-1.35_{-0.07}^{+0.06}$ & $8.32_{-0.02}^{+0.02}$ \\
$8.87_{-0.06}^{+0.06}$ & $20$ & $0.40_{-0.05}^{+0.08}$ & $0.79_{-0.03}^{+0.03}$ & $0.33_{-0.04}^{+0.04}$ & $-0.65_{-0.05}^{+0.05}$ & $1.03_{-0.03}^{+0.03}$ & $-1.51_{-0.05}^{+0.05}$ & $8.20_{-0.02}^{+0.02}$ \\
$9.61_{-0.05}^{+0.06}$ & $26$ & $0.39_{-0.06}^{+0.04}$ & $0.43_{-0.03}^{+0.03}$ & $0.05_{-0.04}^{+0.04}$ & $-0.99_{-0.02}^{+0.02}$ & $0.78_{-0.03}^{+0.04}$ & $-0.87_{-0.04}^{+0.03}$ & $8.47_{-0.01}^{+0.02}$ \\
$9.68_{-0.05}^{+0.05}$ & $22$ & $1.04_{-0.05}^{+0.06}$ & $0.52_{-0.03}^{+0.03}$ & $-0.07_{-0.04}^{+0.04}$ & $-0.92_{-0.10}^{+0.08}$ & $0.90_{-0.03}^{+0.03}$ & $-0.82_{-0.03}^{+0.02}$ & $8.41_{-0.02}^{+0.02}$ \\
\hline\hline
\multicolumn{9}{c}{$2.7 \leq z < 4.0$ in bins of M$_{\ast}$-$\Delta$sSFR} \\
\hline
$8.85_{-0.03}^{+0.04}$ & $27$ & $-0.02_{-0.05}^{+0.06}$ & $0.70_{-0.02}^{+0.02}$ & $0.54_{-0.04}^{+0.04}$ & $-0.44_{-0.09}^{+0.08}$ & $0.91_{-0.02}^{+0.02}$ & $-1.51_{-0.11}^{+0.09}$ & $8.14_{-0.03}^{+0.03}$ \\
$8.78_{-0.05}^{+0.04}$ & $30$ & $0.50_{-0.08}^{+0.09}$ & $0.79_{-0.02}^{+0.01}$ & $0.60_{-0.03}^{+0.03}$ & $-0.48_{-0.04}^{+0.04}$ & $0.98_{-0.01}^{+0.01}$ & $-1.51_{-0.05}^{+0.04}$ & $8.10_{-0.02}^{+0.02}$ \\
$9.47_{-0.04}^{+0.05}$ & $35$ & $0.53_{-0.07}^{+0.05}$ & $0.66_{-0.02}^{+0.02}$ & $0.25_{-0.03}^{+0.03}$ & $-0.91_{-0.08}^{+0.07}$ & $0.93_{-0.02}^{+0.02}$ & $-1.01_{-0.03}^{+0.03}$ & $8.31_{-0.01}^{+0.01}$ \\
$9.33_{-0.04}^{+0.04}$ & $23$ & $1.14_{-0.04}^{+0.04}$ & $0.75_{-0.01}^{+0.01}$ & $0.42_{-0.02}^{+0.02}$ & $-0.65_{-0.03}^{+0.03}$ & $0.97_{-0.01}^{+0.01}$ & $-1.12_{-0.02}^{+0.02}$ & $8.20_{-0.01}^{+0.01}$ \\
\hline\hline
\multicolumn{9}{c}{$4.0 \leq z < 5.0$ in bins of M$_{\ast}$-$\Delta$sSFR} \\
\hline
$8.74_{-0.06}^{+0.05}$ & $19$ & $0.29_{-0.12}^{+0.09}$ & $0.74_{-0.03}^{+0.03}$ & $0.59_{-0.05}^{+0.06}$ & $-0.48_{-0.09}^{+0.08}$ & $0.93_{-0.03}^{+0.03}$ & $-1.55_{-0.16}^{+0.11}$ & $8.12_{-0.04}^{+0.03}$ \\
$8.72_{-0.04}^{+0.05}$ & $15$ & $0.73_{-0.09}^{+0.10}$ & $0.79_{-0.02}^{+0.02}$ & $0.69_{-0.03}^{+0.03}$ & $-0.36_{-0.03}^{+0.03}$ & $0.97_{-0.02}^{+0.02}$ & $-1.51_{-0.05}^{+0.04}$ & $8.03_{-0.02}^{+0.02}$ \\
$9.48_{-0.08}^{+0.06}$ & $21$ & $0.67_{-0.06}^{+0.08}$ & $0.67_{-0.02}^{+0.03}$ & $0.33_{-0.04}^{+0.04}$ & $-0.70_{-0.08}^{+0.06}$ & $0.92_{-0.03}^{+0.03}$ & $-1.05_{-0.04}^{+0.04}$ & $8.26_{-0.02}^{+0.02}$ \\
$9.26_{-0.09}^{+0.09}$ & $13$ & $1.10_{-0.05}^{+0.09}$ & $0.77_{-0.02}^{+0.02}$ & $0.41_{-0.03}^{+0.03}$ & $-0.64_{-0.04}^{+0.04}$ & $0.99_{-0.02}^{+0.02}$ & $-1.05_{-0.03}^{+0.04}$ & $8.19_{-0.01}^{+0.01}$ \\
\hline\hline
\multicolumn{9}{c}{$5.0 \leq z < 7.0$ in bins of M$_{\ast}$-$\Delta$sSFR} \\
\hline
$8.74_{-0.06}^{+0.08}$ & $12$ & $0.28_{-0.09}^{+0.11}$ & $0.72_{-0.03}^{+0.03}$ & $0.80_{-0.06}^{+0.08}$ & $-0.20_{-0.08}^{+0.09}$ & $0.89_{-0.03}^{+0.03}$ & $-1.35_{-0.01}^{+0.01}$ & $7.88_{-0.15}^{+0.09}$ \\
$8.69_{-0.04}^{+0.05}$ & $15$ & $0.81_{-0.04}^{+0.07}$ & $0.78_{-0.02}^{+0.02}$ & $0.77_{-0.04}^{+0.05}$ & $-0.23_{-0.05}^{+0.05}$ & $0.94_{-0.02}^{+0.02}$ & $-1.42_{-0.07}^{+0.06}$ & $7.93_{-0.04}^{+0.04}$ \\
$9.23_{-0.04}^{+0.05}$ & $17$ & $0.76_{-0.09}^{+0.09}$ & $0.75_{-0.02}^{+0.02}$ & $0.47_{-0.03}^{+0.03}$ & $-0.53_{-0.05}^{+0.04}$ & $0.97_{-0.02}^{+0.02}$ & $-1.21_{-0.06}^{+0.05}$ & $8.16_{-0.01}^{+0.02}$ \\
$9.20_{-0.08}^{+0.11}$ & $11$ & $1.24_{-0.07}^{+0.10}$ & $0.79_{-0.02}^{+0.02}$ & $0.59_{-0.03}^{+0.03}$ & $-0.50_{-0.04}^{+0.04}$ & $0.98_{-0.02}^{+0.02}$ & $-1.23_{-0.05}^{+0.04}$ & $8.11_{-0.01}^{+0.02}$ \\
\hline\hline
\enddata
\tablenotetext{a}{Median stellar mass of galaxies in each bin.}
\tablenotetext{b}{Number of galaxies in each bin.}
\tablenotetext{c}{SFR determined from median SFR of constituent galaxies.}
\end{deluxetable*}

%% file: table_mzr_fits_wsanders.tex
\begin{deluxetable}{ccc}[h!t]
\tablecaption{Best-fit MZR for different redshift ranges \label{tab:MZR_fits}}

\tablehead{
    stack &
    $\gamma$\tablenotemark{a} &
    $Z_9$\tablenotemark{b}
}

\startdata
$1.4<z<2.7$ & $0.22 \pm 0.03$ & $8.29 \pm 0.02$ \\
$2.7<z<4.0$ & $0.21 \pm 0.03$ & $8.17 \pm 0.01$ \\
$4.0<z<5.0$ & $0.21 \pm 0.04$ & $8.15 \pm 0.01$ \\
$5.0<z<7.0$ & $0.41 \pm 0.09$ & $8.06 \pm 0.02$ \\
\hline\hline
MOSDEF stack & $\gamma$ & $Z_{10}$ \\\hline
$z\sim2.3$ & $0.21 \pm 0.02$ & $8.45 \pm 0.01$ \\
$z\sim3.3$ & $0.20 \pm 0.04$ & $8.37 \pm 0.02$ \\
\hline\hline
\enddata
\tablenotetext{a}{Slope of the MZR.}
\tablenotetext{b}{Metallicity $12+\log(\mathrm{O/H})$ at $M_\ast=10^9\,M_\odot$.}
\end{deluxetable}